\documentclass[12pt,preprint]{emulateapj}

\begin{document}
\title{$z\sim2$-9 Galaxies magnified by the Hubble Frontier Field Clusters I: Source Selection and Surface Density-Magnification Constraints from $>$2500 galaxies}
\author{R.J. Bouwens\altaffilmark{1}, G. Illingworth\altaffilmark{2}, R.S. Ellis\altaffilmark{3}, P. Oesch\altaffilmark{4,5}, A. Paulino-Afonso\altaffilmark{6}, B. Ribeiro\altaffilmark{1},
  M. Stefanon\altaffilmark{1}}
\altaffiltext{1}{Leiden Observatory, Leiden University, NL-2300 RA Leiden, Netherlands}
\altaffiltext{2}{UCO/Lick Observatory, University of California, Santa Cruz, CA 95064}
\altaffiltext{3}{Department of Physics and Astronomy, University College London, Gower Street, London WC1E 6BT, UK}
\altaffiltext{4}{Department of Astronomy, University of Geneva, Chemin Pegasi 51, 1290 Versoix, Switzerland}
\altaffiltext{5}{Cosmic Dawn Center (DAWN), Niels Bohr Institute, University of Copenhagen, Jagtvej 128, K\o benhavn N, DK-2200, Denmark}
\altaffiltext{6}{CENTRA - Location/Directions, Pavilh{\~ a}o de F{\' i}sica, 4 Piso, Av. Rovisco Pais 1, 1049-001 Lisboa, Portugal}
\begin{abstract}
We assemble a large comprehensive sample of 2534 $z\sim2$, 3, 4, 5, 6,
7, 8, and 9 galaxies lensed by the six clusters from the Hubble
Frontier Fields (HFF) program. Making use of the availability of
multiple independent magnification models for each of the HFF clusters
and alternatively treating one of the models as the “truth,” we show
that the median magnification factors from the v4 parametric models
are typically reliable to values of 30 to 50, and in one case to
100. Using the median magnification factor from the latest v4 models,
we estimate the UV luminosities of the 2534 lensed $z\sim2$-9
galaxies, finding sources as faint as $-$12.4 mag at $z\sim3$ and
$-$12.9 mag at $z\sim7$.  We explicitly demonstrate the power of the
surface density-magnification relations $\Sigma(z)$ vs. $\mu$ in the
HFF clusters to constrain both distant galaxy properties and cluster
lensing properties.  Based on the $\Sigma(z)$ vs. $\mu$ relations, we
show that the median magnification estimates from existing public
models must be reliable predictors of the true magnification $\mu$ to
$\mu<15$ (95\% confidence).  We also use the observed $\Sigma(z)$
vs. $\mu$ relations to derive constraints on the evolution of the
luminosity function faint-end slope from $z\sim7$ to $z\sim2$, showing
that faint-end slope results can be consistent with blank-field
studies if, and only if, the selection efficiency shows no strong
dependence on the magnification factor $\mu$.  This can only be the
case if very low luminosity galaxies are very small, being unresolved
in deep lensing probes.
\end{abstract}

\section{Introduction}

The lowest luminosity galaxies in the early universe represent a key
topic of interest in current studies.  Not only did these sources
likely play a central role in the reionization of the universe (e.g.,
Bunker et al.\ 2004; Kuhlen \& Faucher-Giguere 2012; Robertson et
al.\ 2015; Bouwens et al.\ 2015), but these galaxies are also
potential progenitors to many local stellar systems (Weisz et
al.\ 2014; Boylan-Kolchin et al.\ 2015; Bouwens et al.\ 2017b).  The
lowest luminosity galaxies in the early universe provide us with our
best physical analogues to galaxies at even higher redshifts.
Finally, by probing galaxies in the early universe at $z\sim2$-8, we
can gain insight into the efficiency of star formation in lower-mass
halos at even higher redshifts and provide important constraints for
galaxy formation models.

By combining gravitational lensing from massive galaxy clusters with
long exposures from the {\it Hubble Space Telescope} and other
facilities, the {\it Hubble} Frontier Fields (HFF) program (Coe et
al.\ 2015; Lotz et al.\ 2017) provides us with a powerful means to
probe extremely low luminosity galaxies in the early universe and to
examine their properties.  Already there have been many efforts to
utilize data from the HFF program to find very faint galaxies.  Atek
et al.\ (2014) made early use of the data to probe the prevalence of
$z\sim6$-7 galaxies to $-15$ mag.  Alavi et al.\ (2016) combined
observations available over the HFF clusters, together with Abell 1689
(Alavi et al.\ 2014), to identify plausible candidates to $-12$ mag.
Finally, several groups have identified $z\sim6$-9 candidates with
nominal $M_{UV,AB}$ magnitudes as faint as $\sim-$11 mag (Kawamata et
al.\ 2016; Castellano et al.\ 2016; Livermore et al.\ 2017; Bouwens et
al.\ 2017b; Ishigaki et al.\ 2018; Kawamata et al.\ 2018; Atek et
al.\ 2018; Bhatawdekar et al.\ 2019).

The compilation of such catalogs of faint galaxies has been done with
the goal not only to probe extremely low luminosity galaxies to obtain
insights into their physical properties, but with a goal of
constraining their overall prevalence, quantifying the $UV$-continuum
emissivity of faint galaxies, and determining where if anywhere the LF
might turn over at the faint end.  While constraints on the UV LF have
been presented to $-$15 mag (Atek et al.\ 2015a, 2015b), $-$14 mag
(Ishigaki et al.\ 2018; Kawamata et al.\ 2018; Bhatawdekar et
al.\ 2019), and $-$13 mag (Castellano et al.\ 2016; Bouwens et
al.\ 2017b; Livermore et al.\ 2017; Atek et al. 2018), the faint-end
form of the $UV$ LF has been subject to considerable debate due to
uncertainties in the size distribution and magnification factors for
sources (Bouwens et al.\ 2017; Kawamata et al.\ 2018; Atek et
al.\ 2018; Bouwens et al.\ 2019).  While all recent studies of the
$z\sim6$-7 LFs (Castellano et al.\ 2016; Livermore et al.\ 2017;
Bouwens et al.\ 2017b; Kawamata et al.\ 2018; Ishigaki et al.\ 2018;
Atek et al.\ 2018; Leung et al.\ 2018) agree on the lack of a
turn-over in the $UV$ LF brightward of $-$16 mag, some determinations
are nevertheless consistent with there being a turn-over fainter than
$-$16 mag (Bouwens et al.\ 2017b; Yue et al.\ 2018; Atek et al.\ 2018;
Leung et al.\ 2018).

To improve current efforts to use faint lensed samples to map out the
faint end of the $UV$ LF and redress limitations in previous work, we
conduct a comprehensive study of lensed samples across cosmic time,
from $z\sim9$ to $z\sim2$.  We will consider not only the selection of
galaxies at $z\sim6$-10 and $z\sim2$-3 as done in many studies, but
also demonstrate how $z\sim4$-5 galaxies can be selected from the HFF
data set and used to extend the LF fainter.  By working with samples
over a larger range of redshifts, we will have much more of a baseline
in cosmic time to compare results from blank and lensing cluster
studies, while also obtaining results on the extreme faint end of the
LF.

We will be further pursuing these endeavors in a companion paper
(R. Bouwens et al.\ 2022, in prep).  However, before doing so, we
first focus on the surface density of sources in these samples at
various redshifts and how these surface densities depend on the model
magnification factors.  The relationship between surface density and
magnification has been referred to as the 'magnification bias' (Turner
et al.\ 1984; Broadhurst 1995, Broadhurst et al.\ 2005; Umetsu \&
Broadhurst 2008; Leung et al 2018).  Importantly, the slope of the
surface density vs. magnification relation has been predicted to show
a clear correlation with the faint-end slope of the LF (Broadhurst
1995).  Also impacting the dependence are the sizes of faint sources
as well as the reliability of the magnification models to high values.
Through a careful analysis of the surface density results as a
function of both redshift and magnification factor, we can obtain
critical tests of many of the underlying assumptions essential to
deriving constraints on galaxy LF results from lensed samples.

To maximize the robustness of the luminosities and magnification
factors we estimate for candidate $z=2$-10 sources, we make use of the
full set of recent public lensing models in computing source
luminosities and confidence intervals.  The availability of a larger
set of independently-derived lensing models available for HFF
clusters, taking advantage of an enlarged set of spectroscopic
redshifts (e.g., Owers et al.\ 2011; Schmidt et al.\ 2014; Vanzella et
al.\ 2014; Limousin et al.\ 2016; Jauzac et al.\ 2016; Mahler et
al.\ 2017; Caminha et al.\ 2017), allows for more accurate constraints
on the lensing magnification and uncertainties than could be done
previously.

\begin{figure*}
\epsscale{1.19}
\plotone{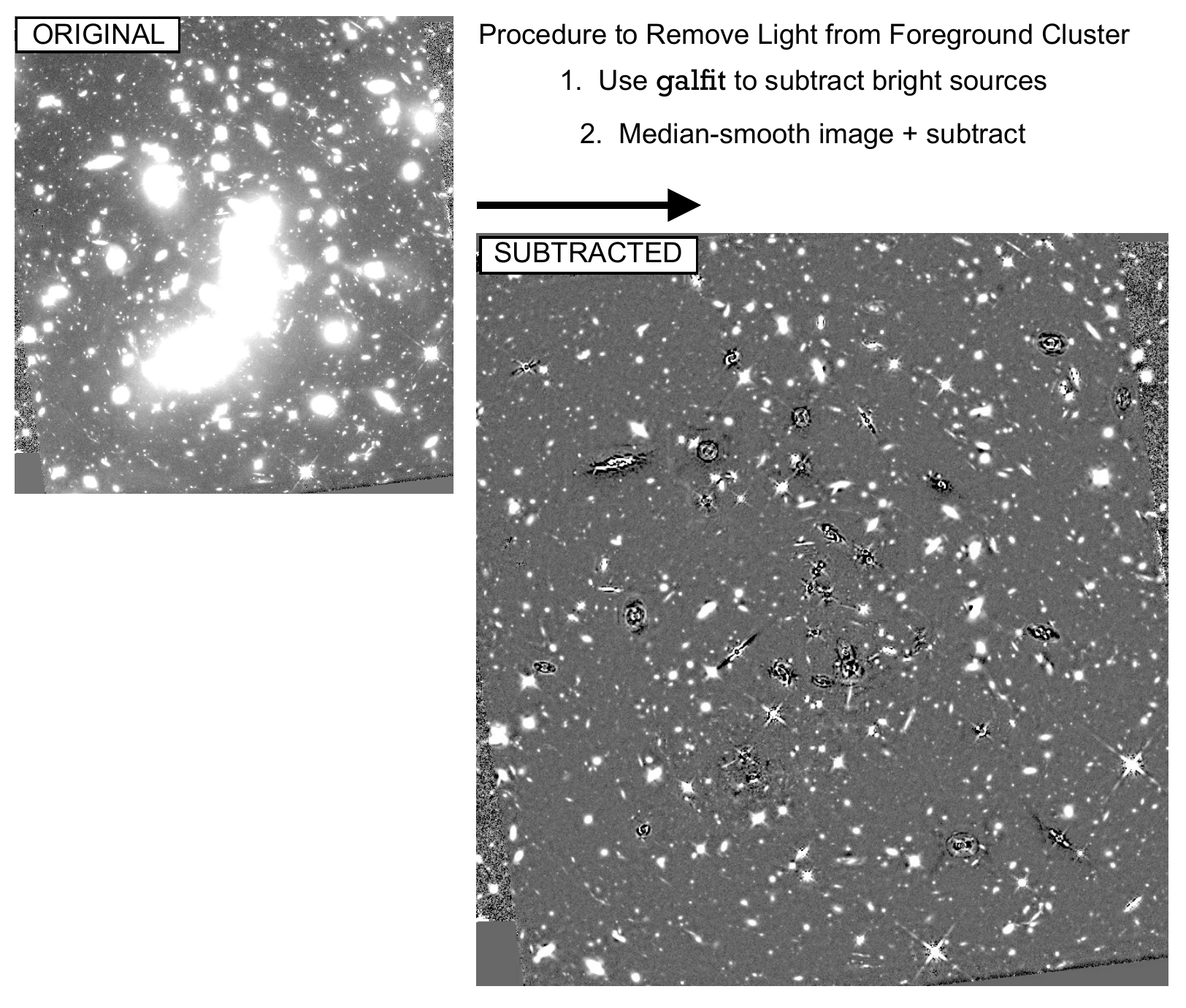}
\caption{Illustration of our procedure to subtract both light from the
  individual cluster galaxies and the cluster (i.e., the intracluster
  light).  Our subtraction procedure (\S2.2) involves two noteworthy
  steps: (1) subtraction of the bright foreground galaxies and (2)
  subtraction of the ICL.  In subtracting bright foreground galaxies,
  we perform first careful profile fits using a S{\'e}rsic
  parameterization and secondly azimuthally-averaged fits to the light
  in various elliptical annuli.  Finally, modeling of the
  intra-cluster light is performed based on two distinct median
  smoothings of the imaging data.  Our intracluster-light subtraction
  procedure performs comparably well to other methodologies employed
  in the literature.\label{fig:illust_iclsub}}
\end{figure*}

The broad organization of this paper is as follows.  We begin by
describing our data set and procedure for subtracting the foreground
light from the cluster (\S2).  We then move onto a description of our
procedure for constructing catalogs, selecting sources, and estimating
the magnification factors (\S3.1-\S3.4), and then we finally compare
our faint samples with previous samples from the literature (\S3.5).
Finally, in \S4, we leverage our large samples to compute the surface
density $\Sigma$ of galaxies at a given redshift vs. magnification
factor $\mu$ and show how this can be used as a constraint on the
faint-end slope of the $UV$ LF at various redshifts, the predictive
power of the lensing models to high magnification factors, and other
issues.  \S5 summarizes our results and provides a prospective for our
future pursuits.  For simplicity, we refer to the {\it HST} F275W,
F336W, F435W, F606W, F814W, F105W, F125W, F140W, and F160W bands as
$UV_{275}$, $U_{336}$, $B_{435}$, $V_{606}$, $I_{814}$, $Y_{105}$,
$J_{125}$, $JH_{140}$, and $H_{160}$, respectively.  Where necessary,
we assume $\Omega_0 = 0.3$, $\Omega_{\Lambda} = 0.7$, and $H_0 =
70\,\textrm{km/s/Mpc}$.  All magnitudes are given in the AB system
(Oke \& Gunn 1983).

\section{Data Sets and Background Subtraction}

\subsection{Data Sets}

The primary data set for this study is the deep ultraviolet, optical,
and near-infrared observations with {\it HST} over the {\it Hubble}
Frontier Fields.  The exposure times of the available {\it HST}
observations vary somewhat from cluster to cluster, but includes at
least 8, 8, 18, 10, 42, 24, 12, 10, and 24 orbits in the $UV_{275}$,
$U_{336}$, $B_{435}$, $V_{606}$, $I_{814}$, $Y_{105}$, $J_{125}$,
$JH_{140}$, and $H_{160}$ bands, respectively.  We made use of the
v1.0 reductions of the HFF provided by Koekemoer et al.\ (2014).  We
reduced the WFC3/UVIS observations using techniques similar to those
used on the HDUV program (Oesch et al.\ 2018b), while ensuring that
the final drizzled product was registered to the optical and near-IR
reduced images from the public v1.0 HFF release.  To account for the
impact of extinction from our Galaxy along the line of sight to the
clusters, we shifted the {\it HST} zeropoints that derived from
Schlafly \& Finkbeiner (2011) using the NASA/IPAC Extragalactic
Database extinction
calculator.\footnote{https://ned.ipac.caltech.edu/extinction\_calculator}

\subsection{Background Subtraction Procedure}

Searches for distant galaxies are made significantly more challenging
due to the light from the foreground galaxy cluster.  This includes
both the intracluster light and light from bright cluster galaxies.
Importantly, these foregrounds cause many interesting sources to be
missed.  This occurs as a result of confusion and foreground light
being mistaken as light blueward of the Lyman break used to select
high-redshift sources.  These effects significantly complicate
searches for distant galaxies.

A wide variety of different procedures have been developed to cope
with the issue of intracluster light, e.g., Oesch et al.\ (2015), Atek
et al.\ (2015b), Merlin et al.\ (2016), Livermore et al.\ (2017),
Shipley et al.\ (2018), and Bhatawdekar et al.\ (2019).  Many groups
have made use of the SExtractor median-filtering algorithm to remove
light from the cluster (Atek et al.\ 2015b), while making use of
\textsc{galfit} (Peng et al.\ 2002, 2010) to subtract light from
foreground cluster galaxies (Merlin et al.\ 2016; Bhatawdekar et
al.\ 2019).  Merlin et al.\ (2016) subtract light from the cluster
adopting a modified Ferrer profile (Binney \& Tremaine 1987) for the
intra-cluster light.  Livermore et al.\ (2017) used a wavelet
procedure to subtract light from the cluster.

\subsubsection{Fits to Individual Cluster Galaxies}

The procedure we utilize involves first fitting to the light profiles
of the 50 brightest foreground galaxies identified in the
3.4$'$$\times$3.4$'$ {\it HST} footprint over the cluster obtained with the
ACS/WFC instrument.

The $I_{814}$-band images provide us with our starting point in
fitting the spatial profile of individual foreground galaxies.
Advantages of the $I_{814}$-band images are (1) their acquisition with
the {\it HST} Advanced Camera for Surveys Wide Field Camera, which features
a narrower PSF ($\sim$0.09$''$ FWHM) than with the WFC3/IR
($\gtrsim$0.15$''$ FWHM) and (2) the $I_{814}$ band's probing the
rest-frame optical light at 0.6$\mu$m in cluster galaxies, but not
probing so red that the intracluster light substantially raises the
background in most clusters.

Fits to the spatial profiles of foreground galaxies are performed with
the \textsc{galfit} software using a S{\'e}rsic profile, allowing the
source center, axial ratio, position angle, source size, S{\'e}rsic
parameter, and total flux to vary.  Our fits are based on a fit to a
2D grid of pixel flux values extending over the segmentation map for
each source and extending 2$''$ further than this in each direction.
After fitting the spatial profile of individual galaxies in the
$I_{814}$ band, we repeat the fits in the other bands, only leaving
the total flux as a free parameter in the fits while fixing the other
S{\'e}rsic parameters to the values from our $I_{814}$ profile fits.

When subtracting these profile fits from the images, we compute the
fit values at the horizontal and vertical boundaries of the
two-dimensional region we are fitting.  Using the fit values at the
boundaries, we add back a linear interpolation of the fit values at
the boundaries to avoid creating a square ``hole'' from the
subtractions.

After using \textsc{galfit} to provide an initial subtraction of light
from individual sources, we refined the subtraction by quantifying the
average surface brightness $S(r_A)$ in various elliptical annuli as a
function of the major-axis radius.  The elliptical annuli are defined
using the position angle and axial ratio derived using our earlier
\textsc{galfit} fits.  The average surface brightness profiles
$S(r_A)$ are then smoothed along the major radial axis, and these
smoothed elliptical annuli are then subtracted from the data.

\begin{deluxetable*}{ccc}
\tablewidth{0cm}
\tabletypesize{\footnotesize}
\tablecaption{Criteria Utilized in Selecting our $z\sim2$-9 Samples\tablenotemark{*}\label{tab:selcrit}}
\tablehead{\colhead{Sample} & \colhead{Criterion} & \colhead{HFF Clusters}}
\startdata
$z$$\sim$2 & (($UV_{275}$$-$$B_{435}$$>$1)$\vee$(($UV_{336}$$-$$V_{606}$$>$1)$\wedge$(SN($UV_{275}$)$<$2))$\wedge$ & Abell 2744, MACS 0416, MACS0717  \\
  &   ($V_{606}$$-$$I_{814}$$<$0.3)$\wedge$(1.5$<$$z_{phot}$$<$2.5)$\wedge$(P($z$$>$1.2)$>$0.65)$\wedge$($\chi^2$$<$25) & MACS 1149, Abell S1063, Abell 370\\\\
$z$$\sim$3 & (($UV_{275}$$-$$B_{435}$$>$1)$\vee$(($UV_{336}$$-$$V_{606}$$>$1)$\wedge$(SN($UV_{275}$)$<$2))$\wedge$ & Abell 2744, MACS 0416, MACS0717  \\
  &   ($V_{606}$$-$$I_{814}$$<$0.3)$\wedge$(2.5$<$$z_{phot}$$<$3.5)$\wedge$(P($z$$>$1.2)$>$0.65)$\wedge$($\chi^2$$<$25) & MACS 1149, Abell S1063, Abell 370\\\\ 
$z$$\sim$4 & $(B_{435}$$-$$V_{606}$$>$1)$\wedge$($I_{814}$$-$$J_{125}$$<$1)$\wedge$$(B_{435}$$-$$V_{606}$$>$1.6($I_{814}$$-$$J_{125}$)+1)$\wedge$ & MACS0717, MACS1149\\
  & ($V_{606}$$-$$I_{814}$$<$0.5)$\wedge$[not in $z\sim5$ selection] & \\\\ 
$z$$\sim$5 & $(V_{606}$$-$$I_{814}$$>$1.2)$\wedge$$(Y_{105}$$-$$H_{160}$$<$0.9)$\wedge$$(V_{606}$$-$$I_{814}$$>$1.2($Y_{105}$$-$$H_{160})$$+$1.32)$\wedge$ & Abell 2744, MACS 0416, Abell S1063  \\
  & (\textrm{SN}($B$)$<$2)$\wedge$[$z\sim5$ non-detection criterion]\tablenotemark{a}$\wedge$[not in $z\sim6$ selection] & Abell 370\\\\ 
$z$$\sim$6 & $(I_{814}$$-$$Y_{105}$$>$0.6)$\wedge$($Y_{105}$$-$$H_{160}$$<$0.45)$\wedge$$(I_{814}$$-$$Y_{105}$$>$0.6($Y_{105}$$-$$H_{160}$))$\wedge$ & Abell 2744, MACS 0416, MACS0717  \\
  & ($Y_{105}$$-$$H_{160}$$<$0.75($J_{125}$$-$$H_{160}$)+0.52)$\wedge$(SN($B_{435}$)$<$2)$\wedge$ & MACS1149, Abell S1063, Abell 370\\
  & ([$z\sim6$ non-detection criterion]\tablenotemark{a}$\vee$($V_{606}$$-$$Y_{105}$$>$2.5))$\wedge$ \\
  & ($z_{phot}<6.3$)$\wedge$(P($z$$>$4.3)$>$0.65)$\wedge$[not in $z\sim 8$ selection] & \\\\ 
$z$$\sim$7 & $(I_{814}$$-$$Y_{105}$$>$0.6)$\wedge$($Y_{105}$$-$$H_{160}$$<$0.45)$\wedge$$(I_{814}$$-$$Y_{105}$$>$0.6($Y_{105}$$-$$H_{160}$))$\wedge$ & Abell 2744, MACS 0416, MACS0717  \\
  & ($Y_{105}$$-$$H_{160}$$<$0.75($J_{125}$$-$$H_{160}$)+0.52)$\wedge$(SN($B_{435}$)$<$2)$\wedge$ & MACS1149, Abell S1063, Abell 370\\
  & ([$z\sim6$ non-detection criterion]\tablenotemark{a}$\vee$($V_{606}$$-$$Y_{105}$$>$2.5))$\wedge$ \\
  & ($z_{phot}>6.3$)$\wedge$(P($z$$>$4.3)$>$0.65)$\wedge$[not in $z\sim 8$ selection] & \\\\ 
$z$$\sim$8 & ($Y_{105}$$-$$J_{125}$$>$0.45)$\wedge$$(J_{125}$$-$$H_{160}$$<$0.5)$\wedge$$(Y_{105}$$-$$J_{125}$$>$0.75($J_{125}$$-$$H_{160}$)$+$0.525)$\wedge$ & Abell 2744, MACS 0416, MACS0717  \\
  & [$z\sim8$ non-detection criterion]\tablenotemark{a} & MACS1149, Abell S1063, Abell 370\\\\
$z$$\sim$9 & $(J_{125}$$-$$H_{160}$$>$1.2)$\wedge$$((H_{160}$$-$[3.6]$<$1.4)$\vee$$(\textrm{SN}([3.6])$$<$2))$\wedge$  & Abell 2744, MACS 0416, MACS0717  \\
 & [$z\sim9$ non-detection criterion]\tablenotemark{a} & MACS1149, Abell S1063, Abell 370
\enddata

\tablenotetext{*}{Throughout this table, $\wedge$ and $\vee$ represent
  the logical \textbf{AND} and \textbf{OR} symbols, respectively, and
  SN represents the signal to noise.  The $\chi^2$ statistic is as
  defined in Bouwens et al.\ (2011).  In the application of these
  criteria, flux in the dropout band is set equal to the $1\sigma$
  upper limit in cases of a non-detection.}

\tablenotetext{a}{Non-detection criteria for our $z\sim$6 and $z\sim7$
  selection are as follows: $(\chi_{opt} ^2(B_{435},V_{606})<2)$.  For
  our $z\sim 8$ and $z\sim9$ selection, the criteria are the
  following: $(\chi_{opt,0.35''} ^2 < 4)\wedge (\chi_{opt,Kron} ^2 <
  4)\wedge(\chi_{opt,0.2''} ^2 < 4)$ where $\chi^2$ is as defined in
  Bouwens et al.\ (2011), $opt$ includes the $B_{435}$, $V_{606}$, and
  $I_{814}$ bands, and $Kron$ indicates the small scalable apertures
  discussed in \S3.1.}
\end{deluxetable*}

\begin{figure*}
\epsscale{1.17}
\plotone{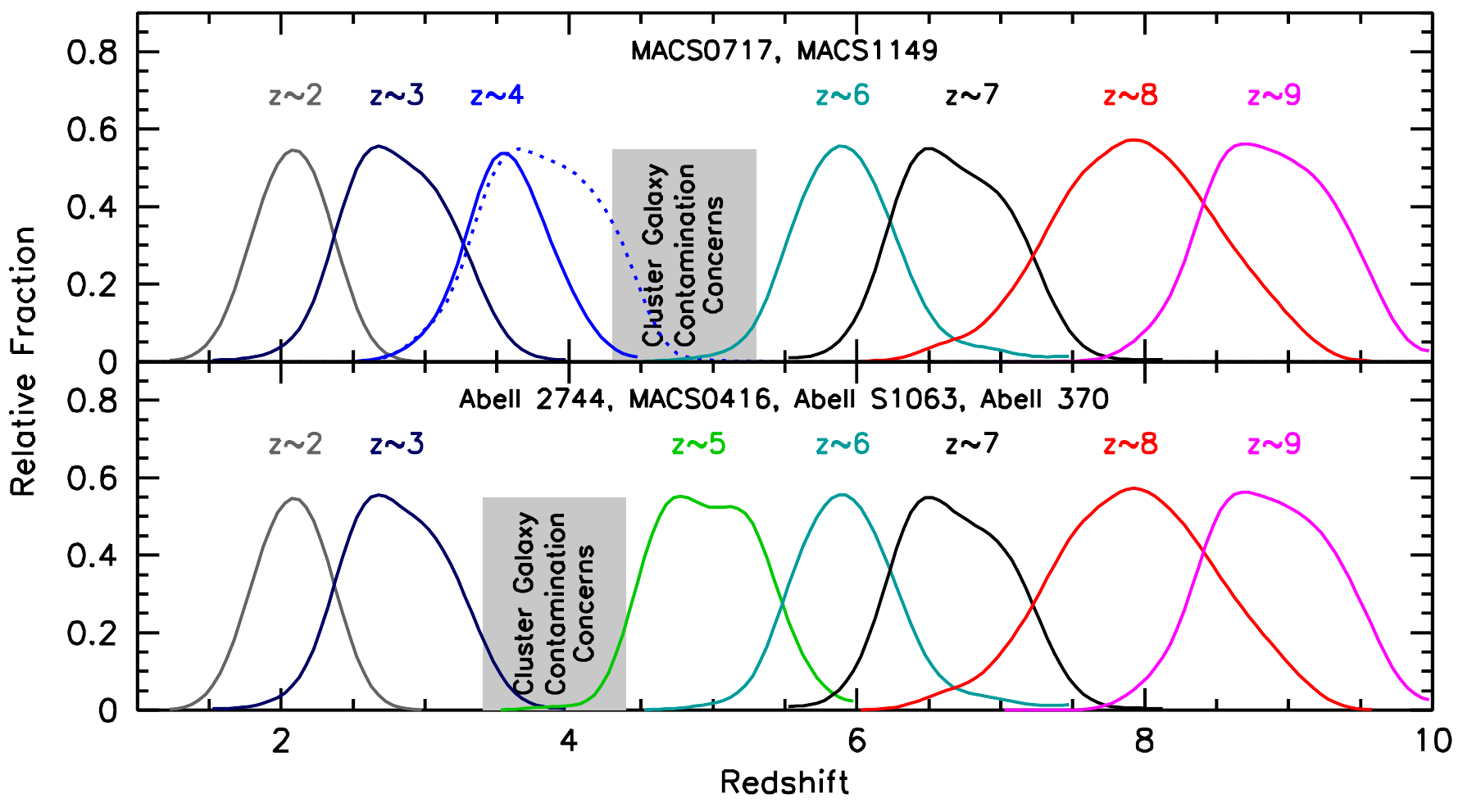}
\caption{The expected redshift distributions of our $z\sim2$, 3, 4, 5,
  6, 7, 8, and 9 selections over the HFF clusters for the two higher
  redshift ($z\sim0.55$) clusters from the HFF program (MACS0717 and
  MACS1149: \textit{upper panel}) and four HFF clusters at lower
  redshifts ($z\sim0.3$-0.4: Abell 2744, MACS0416, Abell S1063, and
  Abell 370).  These redshift distributions are computed from source
  injection and recovery experiments (as e.g. Bouwens et al.\ 2021a
  utilize).  Due to the similar wavelength of the Lyman break in
  $z\sim4$ and $z\sim5$ galaxies to the Balmer/4000\AA$\,$ break in
  foreground galaxies that make up the lower redshift and higher
  redshift clusters (see Figure~\ref{fig:clust_contam}), we only
  consider selections of $z\sim4$ galaxies for the two higher redshift
  clusters and only consider selections of $z\sim5$ galaxies for the
  four lower redshift clusters.  The dotted blue line in the upper
  panel shows the expected $z\sim4$ redshift distribution when
  utilizing our normal $z\sim4$ selection criteria (which would be
  appropriate if a foreground cluster were not present in our search
  field).\label{fig:zdist}}
\end{figure*}

\subsubsection{Modeling of the Intracluster Light}

After having modeled and subtracted the light from the brightest 50
foreground galaxies, the next step we utilize is to remove the
extended intracluster light by subtracting a median-smoothed image of
the cluster -- which we implement in two steps.  Our approach follows
the SExtractor (Bertin \& Arnouts 1996) implementation of this
algorithm.  The procedure is first to break up the image into grid
cells of some fixed individual dimension and set the nominal
background level for a cell to be 3 times the mean minus two times the
median.  Then, the next step is to execute a 3$\times$3 median
smoothing on this grid and construct an ICL model by splining over
individual cells in the grid.  Finally, this ICL model is subtracted
from the data itself.

In our initial implementation of the subtraction of this
median-smoothed image, we adopted a relatively large angular size for
individual cells in the grid, i.e., 1.2$''$, since we found that had
the least impact on the total flux measurements we made for individual
sources.  We, however, found that this did not produce a flat
background across the full image.  In particular, around bright
galaxies, there was some noticeable ``ringing'' in the background,
making the recovery of sources in the affected regions more
challenging.

To improve the flattening of the background in the specific regions of
the image impacted by such ringing, we explicitly demarcated the
affected regions with \textsc{ds9}.  We repeated our determination of
a median-smoothed image using significantly smaller grid cells, i.e.,
with dimension 0.4$''$, and repeated our construction of a
median-smoothed image for the cluster.  Not surprisingly, after
subtraction of this high-resolution median-smoothed image from the
original, much less ringing is evident around bright foreground
galaxies.  This approach, however, results in the systematic reduction
(by $\sim$0.1-0.15 mag) in the measured flux of sources, due to the
subtraction of flux on larger spatial scales.

To maximize the completeness of the present search results around
bright sources while maximizing the accuracy of the photometry in more
crowded regions, we make use of two different median-smoothed images
to subtract the ICL light: (1) using the image with the coarser grid
cells in the less crowded regions, which constitute 80\% of the HFF
WFC3/IR search area, and (2) using the image with finer grid cells in
the more crowded region, constituting 20\% of the image.

\subsubsection{Summary}

The ICL subtraction that the above procedure delivers is illustrated
in Figure~ \ref{fig:illust_iclsub}, and it is clear that we obtain a
good subtraction of the foreground light from the cluster and under
the brightest sources in the cluster, without substantial ``ringing''
in the immediate vicinity of various foreground sources.

In Bouwens et al.\ (2017b: Appendix A), we already demonstrated that
with our background subtraction algorithm (now described in detail
here) we were able to identify a comparable number of $z\sim6$-8
galaxies to that found in Merlin et al. (2016) and Livermore et
al.\ (2017).  The source numbers presented in \S\ref{sec:finalsamples}
provide further evidence that our algorithm works well.

\begin{figure}
\epsscale{1.17}
\plotone{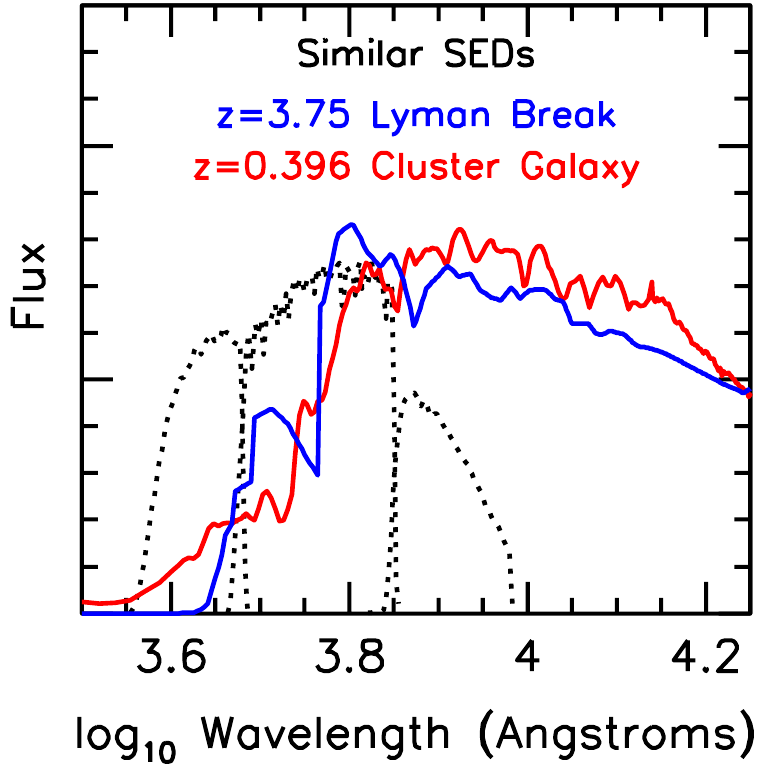}
\caption{Illustration of how similar the SED shape of a $z\sim3.75$
  star-forming galaxy (\textit{blue}) is to that of an evolved galaxy
  at $z=0.396$ (\textit{red}).  The dotted black lines show the
  transmission curves of the optical $B_{435}$, $V_{606}$, and
  $I_{814}$ filters utilized for the six clusters in the HFF program.
  As a result of the similar SED shapes of $z\sim4$ Lyman-break
  galaxies to evolved galaxies in the foreground clusters,
  contamination poses a significant challenge for $z\sim4$ selections
  from the HFF program.\label{fig:clust_contam}}
\end{figure}

\begin{figure}
\epsscale{1.17}
\plotone{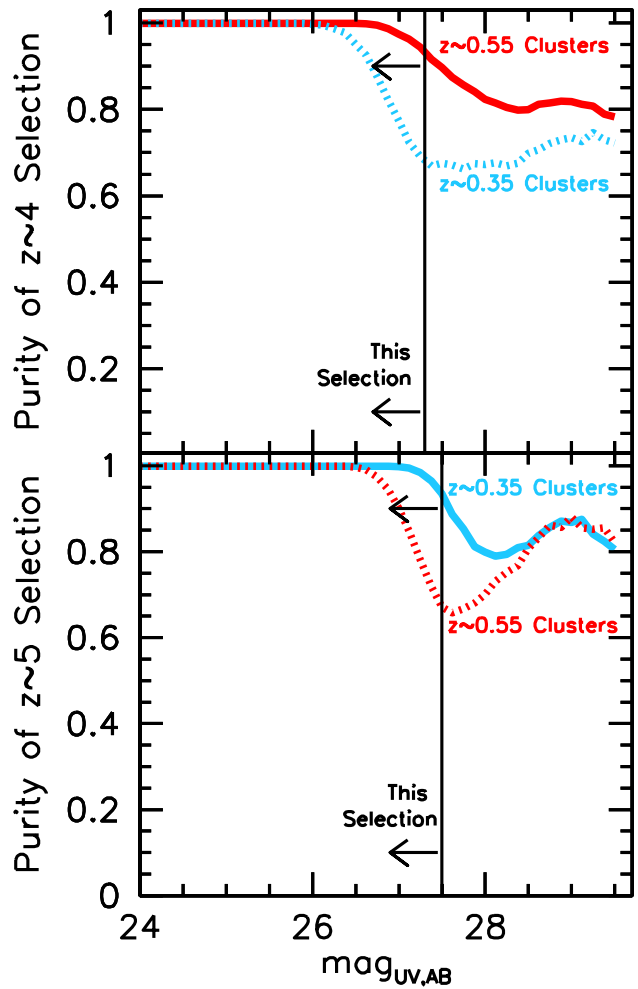}
\caption{Illustration of the estimated purity of our $z\sim4$ and
  $z\sim5$ galaxy selections (\textit{upper and lower panels,
    respectively}) as a function of apparent magnitude behind the HFF
  clusters.  Results are shown both using the $z\sim0.35$ and
  $z\sim0.55$ HFF clusters (\textit{blue and red lines,
    respectively}).  See \S\ref{sec:z45} for a description of the
  simulations.  Despite the challenges created by foreground cluster
  galaxies contaminating selections of star-forming galaxies at
  $z\sim4$-5, relatively clean selections are possible if (1) only
  sources brighter than 27.3 mag at $z\sim4$ and 27.5 mag at $z\sim5$
  (\textit{vertical lines}) are included and (2) only utilizing those
  clusters where the Lyman break occurs across a different set of
  broadband filters than the 4000\AA/Balmer break. \label{fig:purity}}
\end{figure}

\section{Construction of Faint Samples}

\subsection{Catalog Construction}

As in previous work (Bouwens et al.\ 2011, 2015), we use the
SExtractor software (Bertin \& Arnouts 1996) to handle source
detection and photometry.  SExtractor is run in dual-image mode, with
the detection image taken to equal the square root of $\chi^2$ image
(Szalay et al.\ 1999: similar to a coadded image) constructed from one
or more {\it HST} bands, depending on the redshift of galaxy we are
searching.  We use the $B_{435}V_{606}I_{814}$ images for our $z\sim2$
and $z\sim3$ selections, the $Y_{105}$, $J_{125}$, $JH_{140}$ and
$H_{160}$ images for our $z\sim4$, $z\sim5$, $z\sim6$, and $z\sim7$
selections, the $J_{125}$, $JH_{140}$, and $H_{160}$ images for our
$z\sim8$ selection, and the $JH_{140}$ and $H_{160}$ bands for our
$z\sim9$ selection.  Color measurements are made in small scalable
apertures (Kron [1980] factor of 1.2), after PSF-matching the
observations.  The PSF matching is done to the $I_{814}$ band (if the
color consists entirely of optical bands) or the $H_{160}$ band (if
the color also includes a near-infrared band).

Total-magnitude measurements are made by correcting the
smaller-scalable aperture flux measurements to account for light
outside these apertures.  Two corrections are considered.  The first
accounts for the excess flux measured in a larger-scalable aperture
(Kron [1980] factor of 2.5) relative to the smaller-scalable aperture
and second to account for the light on the wings on the PSF (typically
a $\sim$0.15-0.25 mag correction).  The second correction is made
using the tabulated encircled energy distributions (Dressel et
al. 2012).

In selecting $z\sim9$ galaxies over our fields, it is useful to be
able to make use of the sensitive {\it Spitzer}/IRAC observations that
have been obtained over the fields.  Due to the relatively broad PSF
of {\it Spitzer}/IRAC relative to the surface density of sources on
the sky, source crowding is an issue in measuring fluxes for sources.
As in much of our previous work, we use the \textsc{mophongo} software
for this endeavor (Labb{\'e} et al.\ 2010a, 2010b, 2013, 2015).  In
executing this activity, \text{Mophongo} considers the positions and
morphologies of sources in the {\it HST} data as well as an {\it
  HST}-to-IRAC PSF matching kernel to model the spatial distribution
of light from each source.  After fitting for the flux of each source,
light from neighboring sources is subtracted, and then aperture
photometry is performed to measure the flux from a source within a
given aperture.  That measured flux is then corrected to total based
on the form of the PSF.

\subsection{Source Selection\label{sec:selection}}

In selecting $z=2$-9 galaxies from the parallel and cluster fields
that make up the HFF program, we pursue a very similar approach as to
what we have used in Bouwens et al.\ (2021a) over the HDUV fields and
HFF fields.  A summary of the selection criteria we utilize is
presented in Table~\ref{tab:selcrit} for our $z\sim2$, 3, 4, 5, 6, 7,
8, and 9 samples.  

Figure~\ref{fig:zdist} shows the expected redshift distributions of
our $z\sim2$, 3, 4, 5, 6, 7, 8, and 9 selections for the six HFF
clusters considered here.  The expected redshift distributions are as
derived in Bouwens et al.\ (2021a) and make use of the same
$UV$-continuum slope $\beta$ and size-luminosity scalings as
implemented there, with the exception of $z\sim4$ as discussed below.

The best-fit photometric redshifts and $\chi^2$ we derive that play a
role in our $z\sim2$-3 and $z\sim6$-7 selections use the EAzY
photometric redshift software (Brammer et al.\ 2008).  The SED
templates utilized with EAzY were the EAzY\_v1.0 set together with SED
templates from the Galaxy Evolutionary Synthesis Models (GALEV:
Kotulla et al.\ 2009).  Emission lines were added to the later
templates assuming a rest-frame EW of 1300\AA$\,$for H$\alpha$ with
line ratios given by the $0.2 Z_{\odot}$ Anders \&
Fritze-v. Alvensleben (2003) prescription.  When deriving photometric
redshift constraints and computing a $\chi^2$, an additional 7\%
uncertainty in our flux measurements is assumed to allow for small
systematic differences between the observed and model SEDs and small
systematics in the photometry.

Candidate galaxies at $z\sim6$-7, $z\sim8$, and $z\sim9$ are required
to have a signal-to-noise ratio of 6 in a stack of the available
WFC3/IR observations in the $Y_{105}J_{125}JH_{140}H_{160}$,
$J_{125}JH_{140}H_{160}$, and $JH_{140}H_{160}$ bands, respectively.
Sources which correspond to diffraction spikes are the clear result of
an elevated background around a bright source (e.g., for a bright
elliptical galaxy), or correspond to other artifacts in the data are
removed by visual inspection.

All bright ($H_{160,AB}<27$) sources with SExtractor stellarity
parameters in excess of 0.9 (where 0 and 1 correspond to extended and
point sources, respectively) are removed.  Sources where the
stellarity parameter is in excess of 0.6 are also removed, if the {\it
  HST} photometry is much better fit with SEDs of low-mass stars
($\Delta \chi^2 > 2$) from the SpeX library (Burgasser et al.\ 2004)
than with a linear combination of galaxy templates from EAzY (Brammer
et al.\ 2008).

To ensure our $z\sim9$ sample is not contaminated by lower-redshift
sources with red spectral slopes, we made use of the available
3.6$\mu$m {\it Spitzer}/IRAC observations.  Sources were excluded from
our selection if they were detected at $3\sigma$ in the 3.6$\mu$m
observations and had $H_{160}-[3.6]$ colors redder than 0.8 mag.

To maximize the completeness in selecting sources for our $z\sim 6$-9
catalogs (where the numbers are small), we considered five different
background subtraction algorithms, each using a different angular size
for the SExtractor background mesh, and regenerated our source
catalogs and $z\sim 6$-9 selections from each.  This reduces the
number of sources that would be missed due to the deblending choices
made by SExtractor in constructing a given catalog, thereby maximizing
the robustness of our results.

\begin{deluxetable}{ccc}
\tablecolumns{3}
\tablewidth{0cm}
\tabletypesize{\footnotesize}
\tablecaption{Lensing Models Utilized (see also \S\ref{sec:models})\tablenotemark{a}\label{tab:models}}
\tablehead{\colhead{Cluster} & \colhead{Model} & \colhead{Version}}
\startdata
Abell 2744 & CATS (P) & v4.1 \\
           & Sharon/Johnson (P) & v4 \\
           & Keeton (P) & v4 \\
           & GLAFIC (P) & v4 \\
           & Zitrin/NFW (P) & v3 \\
           & Grale (NP) & v4 \\
           & Bradac (NP) & v2 \\
           & Zitrim-LTM-Gauss (NP) & v3 \\
           & Diego (NP) & v4.1 \\\\
MACS0416 & CATS (P) & v4.1 \\
         & Sharon/Johnson (P) & v4 \\
         & Keeton (P) & v4 \\
         & GLAFIC (P) & v4 \\
         & Zitrin/NFW (P) & v3 \\
         & Caminha (P) & v4 \\
         & Grale (NP) & v4 \\
         & Bradac (NP) & v3 \\
         & Zitrim-LTM-Gauss (NP) & v3 \\
         & Diego (NP) & v4.1 \\\\
MACS0717 & CATS (P) & v4.1 \\
         & Sharon/Johnson (P) & v4 \\
         & Keeton (P) & v4 \\
         & GLAFIC (P) & v3 \\       
         & Grale (NP) & v4.1 \\
         & Diego (NP) & v4.1 \\\\
MACS1149 & CATS (P) & v4.1 \\ 
         & Sharon/Johnson (P) & v4 \\ 
         & Keeton (P) & v4 \\ 
         & GLAFIC (P) & v3 \\
         & Grale (NP) & v4 \\
         & Diego (NP) & v4.1 \\\\
Abell 370 & CATS (P) & v4 \\ 
          & Sharon/Johnson (P) & v4 \\ 
          & Keeton (P) & v4 \\ 
          & GLAFIC (P) & v4 \\
          & Bradac (NP) & v4.1 \\
          & Grale (NP) & v4 \\
          & Diego (NP) & v4.1 \\\\
Abell S1063 & CATS (P) & v4.1 \\ 
            & Sharon/Johnson (P) & v4 \\ 
            & Keeton (P) & v4 \\ 
            & GLAFIC (P) & v4 \\
            & Caminha (P) & v4\tablenotemark{c} \\
            & Grale (NP) & v4 \\
            & Diego (NP) & v4.1
\enddata 
\tablenotetext{a}{This includes all publicly available lensing models
  which have high-resolution mass maps for all six HFF clusters.}
\tablenotetext{b}{Parametric models assume that mass in clusters is
  in the form of one or more dark matter components with an
  ellipsoidal Navarro-Frenk-White (NFW: Navarro et al.\ 1997) form and
  include a contribution from galaxies following specific
  mass-to-light scalings.  Two well-known parametric modeling codes
  are \textsc{lenstool} (Jullo \& Kneib 2009) and \textsc{GLAFIC}
  (Oguri 2010).  For the non-parametric models, both assumptions are
  typically relaxed, and the mass distributions considered  typically allow for much more flexibility than with the
  parametric models.}
\tablenotetext{c}{The parametric lens model from Caminha et
  al.\ (2016) for Abell S1063 was kindly made available to the authors
  and makes use of similar constraints to the v4 models.  Thus far, it
  has not been made publicly available on the HFF web site.}
\end{deluxetable}

\subsection{Selection of Galaxies at $z\sim4$-5\label{sec:z45}}

Selecting galaxies at $z\sim4$-5 behind the galaxy clusters in the HFF
program is significantly more challenging than at $z\gtrsim 6$ and
$z\sim 2$-3 due to the cluster galaxies themselves.  Many cluster
galaxies show prominent 4000\AA$\,$ and Balmer breaks which fall
between the $B_{435}$ and $V_{606}$ bands or in the center of the
$V_{606}$ band.  As Figure~\ref{fig:clust_contam} illustrates, this
feature can look very similar to the Lyman break for a $z\sim4$
galaxy.  This makes the selection of $z\sim4$ galaxies especially
challenging for the four lowest-redshift clusters in the HFF program
lying between $z\sim0.3$ and $z\sim0.4$, i.e., Abell 2744, MACS0416,
Abell S1063, and Abell 370.

Given these challenges, we select $z\sim4$ galaxies by targeting the
two clusters where the 4000\AA$\,$ or Balmer breaks are redshifted
significantly through the $V_{606}$ band, i.e., MACS0717 ($z=0.543$)
and MACS1149 ($z=0.547$), such that the 4000\AA/Balmer spectral break
in cluster galaxies is approximately midway through the $V_{606}$
band.  Given that cluster galaxies have $V_{606}-I_{814}$ and
$B_{435}-V_{606}$ colors somewhat similar to $z\sim5$ galaxies, one
can achieve the most robust selection of $z\sim4$ galaxies by
targeting galaxies over the range $z\sim3.5$-4.0, with smaller
$V_{606}-I_{814}$ breaks, i.e., where $(V_{606}-I_{814}<0.5)$.

By contrast, $z\sim5$ galaxies can be best selected by focusing on the
four clusters where the 4000\AA$\,$ or Balmer breaks are not as
significantly redshifted through the $V_{606}$ band, i.e., Abell 2744
($z=0.308$), MACS 0416 ($z=0.396$), Abell S1063 ($z=0.348$), and Abell
370 ($z=0.375$).  The $z\sim5$ criteria we utilize
(Table~\ref{tab:selcrit}) are identical to those presented in Bouwens
et al.\ (2021a).

To minimize the impact of noise in causing sources from the foreground
clusters to satisfy our $z\sim4$-5 selection criteria, we only include
$z\sim4$ and $z\sim5$ sources brighter than $H_{160,AB}\sim 27.3$ mag
and 27.5 mag, respectively, in our selection.  We arrived at these
magnitude limits using Monte-Carlo simulations where we added
photometric noise to the $z\sim0$ Coleman et al.\ (1980) E/S0, Sab,
and Imm templates placed at the approximate redshifts of the HFF
clusters and a 100 Myr Bruzual \& Charlot (2003) starburst template
placed at $z\sim3.75$ and $z\sim5$.  Then, to estimate an approximate
purity at a given magnitude level, we divided the estimated number of
high-redshift galaxies by that number plus the expected contamination
from cluster galaxies.  For these estimates, we assume the surface
density of foreground cluster galaxies is 10$\times$ higher than the
high-redshift population to be conservative.

The result is shown in Figure~\ref{fig:purity} for both our $z\sim4$
and $z\sim5$, and it is clear that we can obtain a much cleaner
selection of $z\sim4$ and 5 galaxies if we restrict ourselves to
galaxies brighter than 27.3 mag and 27.5 mag, respectively.  As a
check on our decision to select $z\sim4$ galaxies from the two highest
redshift clusters, i.e., MACS0717 and MACS1149, and $z\sim5$ galaxies
from the four lower redshift clusters, we repeated these simulations
for the clusters excluded from our $z\sim4$ and $z\sim5$ selections
(i.e., where the Lyman break for $z\sim4$-5 galaxies lies across a
similar set of bands to the 4000\AA/Balmer breaks for cluster
galaxies).  We see significant contamination setting in $\sim$1 mag
brighter for these clusters and being 2-10$\times$ higher.  Based on
these results, it clearly seems to have been prudent to have excluded
these specific cluster fields from our $z\sim4$ and $z\sim5$
selections.

\begin{figure*}
\epsscale{0.97}
\plotone{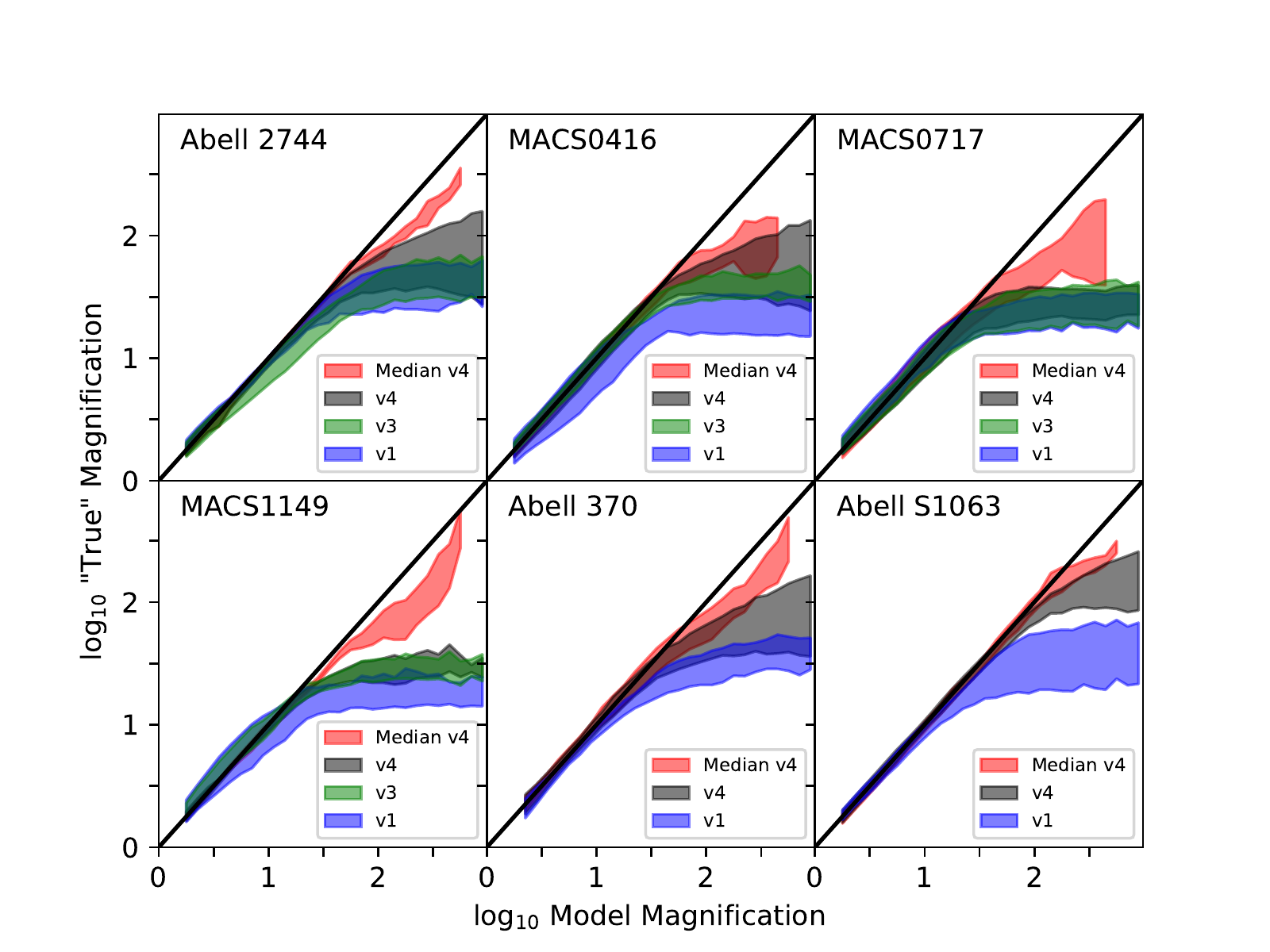}
\caption{Illustration of how well the model magnification factors from
  individual parametric lensing models (\textit{shown along horizontal
    axis}) are able to predict the magnification factors
  (\textit{shown on vertical axis}) for the v4 CATS, Sharon/Johnson,
  Keeton, and \textsc{GLAFIC} models of the same HFF cluster (see
  \S\ref{sec:testing}).  The blue, green, and black-shaded regions
  show the $1\sigma$ range in mean magnification factors predicted
  using independent v1, v3, and v4 models, respectively.  All sources
  are assumed to be at $z=6$.  The red-shaded region shows the
  $1\sigma$ range in mean magnification factors predicted using the
  median of the other v4 parametric models (except the v4 model being
  predicted).  These shaded regions demonstrate the individual
  magnification models are effective in predicting the magnification
  factors to values of $\sim$20-50, depending on the HFF cluster.  In
  general, the v4 models exhibit better predictive power than the v1
  models, particularly for MACS0416, MACS1149, Abell 370, and Abell
  S1063.  The median v4 models exhibit even better predictive power,
  being effective in predicting the magnification factors to values of
  $\sim$100 and in some cases higher.\label{fig:predictind}}
\end{figure*}

\subsection{Model Magnification Factors\label{sec:models}}

To interpret the faint high-redshift sources we have identified behind
the HFF lensing clusters, we must estimate the degree to which each
source is magnified by the foreground cluster.  

Fortunately, in parallel with the acquisition of deep {\it HST}
observations from the HFF program, a variety of groups have been
involved in constructing lensing models for the HFF clusters (e.g.,
Bradac et al. 2009; Diego et al. 2005, 2007, 2015a, 2015b, 2018; ,
Keeton 2010; Liesenborgs et al. 2006; Mahler et al. 2018; Merten et
al. 2015, Richard et al.\ 2014; Zitrin et al. 2012, 2015).  At
present, all six clusters have anywhere between 6 to 10 models
available for them (Table~\ref{tab:models}) using the latest (v3/v4)
sets of constraints.  These constraints include large numbers of
multiple image pairs available from the HFF observations, as well as
spectroscopic redshift constraints mentioned earlier.

Magnification of a source at a particular redshift can be calculated
on the basis of the convergence $\kappa$ and shear $\gamma$ maps
numerous teams have derived in modeling background sources in the HFF
data:
\begin{equation}
\mu = \frac{1}{|(1-\kappa)^2 - \gamma^2)|}
\end{equation}
Note that the $\gamma$ and $\kappa$ values in the above equation are
not simply the values in the published maps, but also include
multiplication by a $D_{LS}/D_{S}$ factor, where $D_{LS}$ is the
angular-diameter distance from the lens to the lensed background
source, while $D_{S}$ is the angular-diameter distance to the lensed
background source.

\subsubsection{Testing the Reliability of the Magnification Factors from the Public Models\label{sec:testing}}

Prior to making use of the public magnification models, it is useful
to ask ourselves over which range of magnification factors these
models can be trusted to give reliable results.  There have been
several earlier studies that have looked into this in some detail
(e.g., Meneghetti et al.\ 2017; Priewe et al.\ 2017; Bouwens et
al.\ 2017a), and one consensus conclusion from these studies has been
that the parametric magnification models show excellent predictive
power to at least a magnification factor of 30.

Following the release of v3 magnification models, more spectroscopic
redshift measurements on lensed galaxies behind the HFF clusters have
been made available.  The release of the v4 lensing models should
provide even more accurate constraints on the magnification factor for
individual sources than the v3 models.  It therefore makes sense to
reassess how well various public lensing models predict the true
magnification factors of sources behind the HFF clusters.

\begin{figure*}
\epsscale{0.97}
\plotone{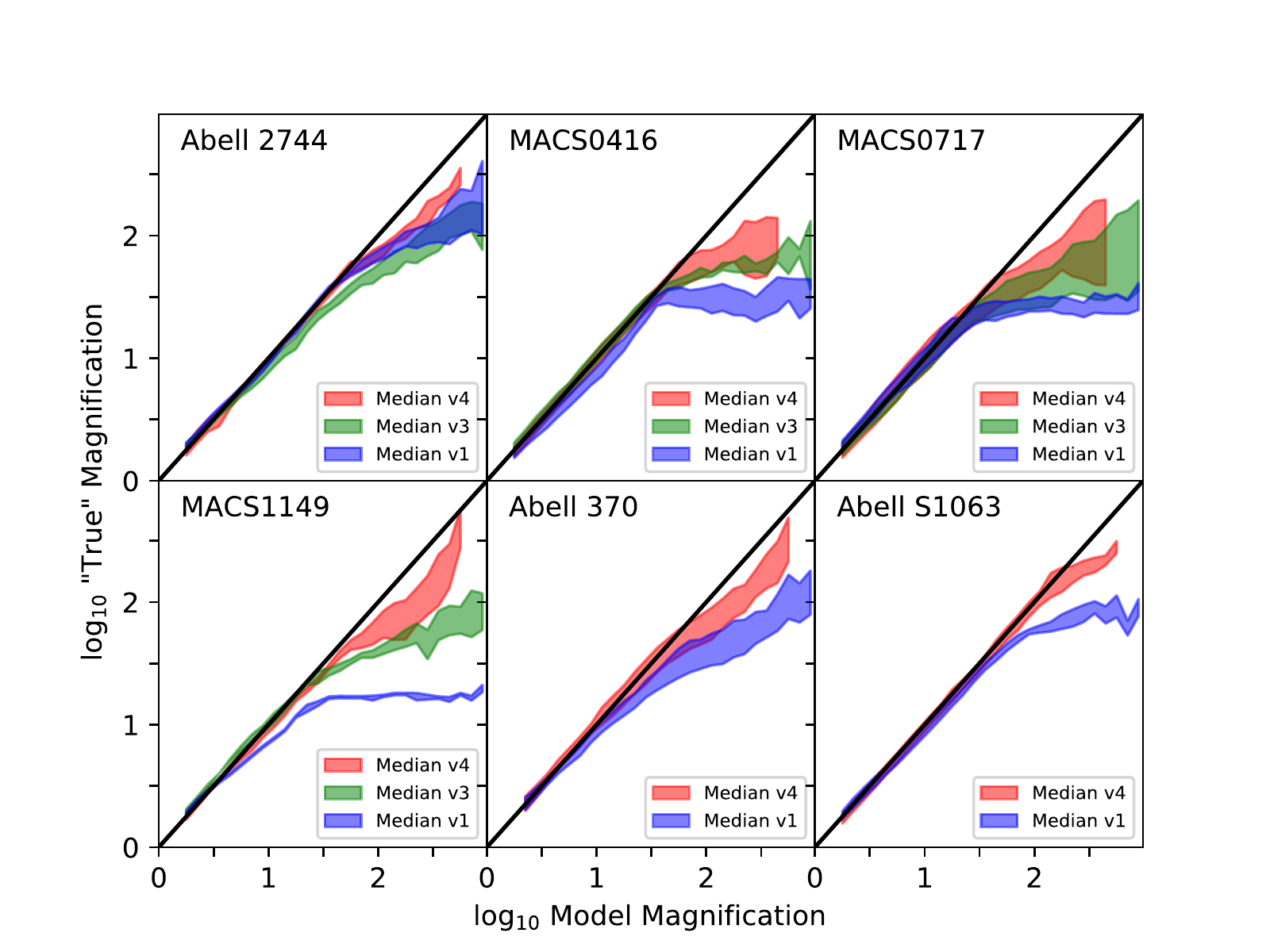}
\caption{An illustration of how well the median of the v1, v3, and v4
  parametric magnification models are able to predict the ``true''
  magnification factor (as represented by one of the v4 parametric
  models: see \S\ref{sec:testing}).  The blue, green, and red shaded
  regions show the $1\sigma$ range of mean magnification factors
  recovered for v4 models at a given median magnification factor from
  the v1, v3, and v4 parametric models, respectively.  All sources are
  assumed to be at $z=6$.  There is a clear improvement in the
  predictive power of the median models from v1 to
  v4.\label{fig:predict}}
\end{figure*}

Leveraging the approach presented in \S3.1 of Bouwens et al.\ (2017a),
we test the predictive power of the lensing model by treating one of
the many parametric lensing models as the truth and then testing how
well the other lensing models are able to predict its magnification
factors.  As each model represents a comparably realistic
representation of the true lensing model, this is a reasonable way to
proceed.  Given the large number of public parametric models available
for each of the individual HFF clusters, we can make use of a
significant number of model pairs to assess the predictive power of
individual models.

In Figure~\ref{fig:predictind}, we present the results we obtain of
this exercise using the magnification factors from individual v1, v3,
and v4 models to predict the magnification factors of separate v4
models for all HFF clusters.  For simplicity, all sources are assumed
to have a redshift of $z=6$ through this exercise.  To avoid favoring
any one of the public lensing models, each family of the v4 lensing
models is alternatively treated as the truth, and then the mean
logarithmic magnification factor in the ``truth'' model is calculated
as a function of the magnification factor for a different lensing
model.  Making use of the many different pairs of lensing models, one
treated as the ``truth'' and the other treated as a predictive model,
we can calculate a mean magnification and $1\sigma$ scatter about the
``true'' magnification, as a function of the model magnification
factor.

\begin{figure*}
\epsscale{0.98} \plotone{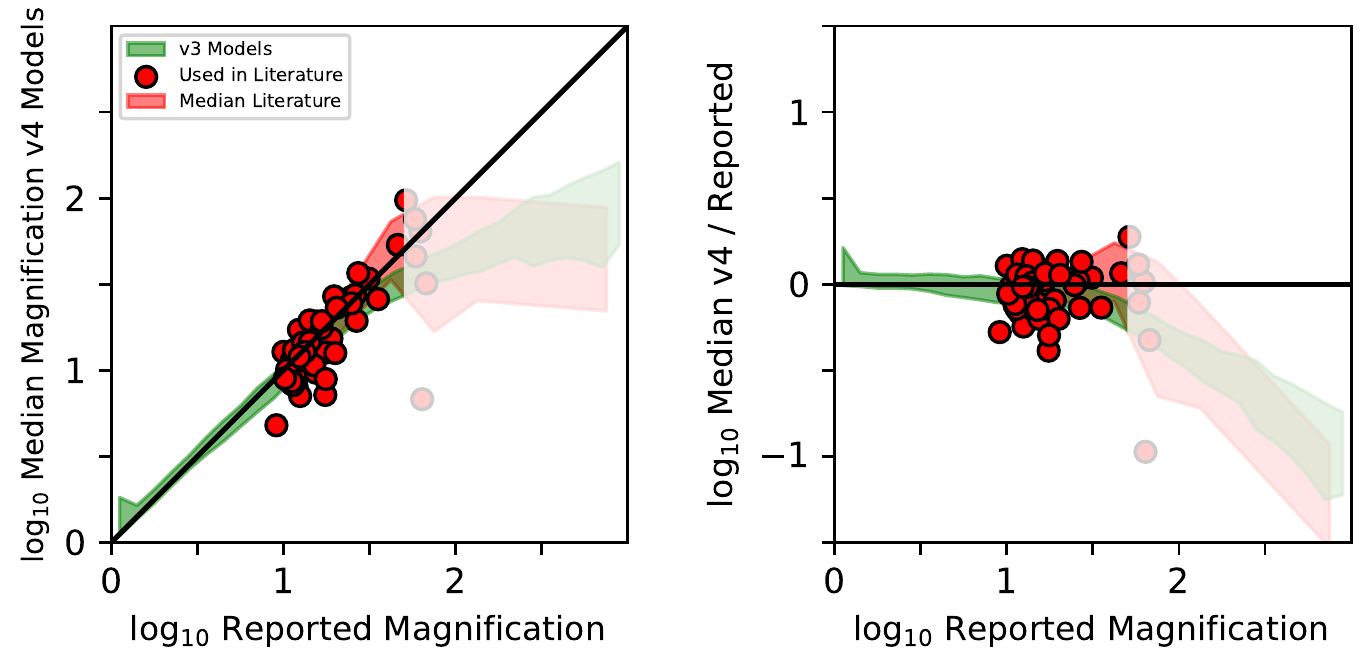}
\caption{(\textit{left}) Comparison of the magnification factors
  reported in the literature (Zitrin et al.\ 2014; Infante et
  al.\ 2015; Livermore et al.\ 2017; Ishigaki et al.\ 2018:
  \textit{red circles}) for specific $z\sim6$-10 galaxies to values
  $<$100 (\textit{horizontal axis}) and median magnification factors
  from the v4 parametric models (see \S\ref{sec:testing}).  The green
  shaded region shows the expected $1\sigma$ range in predictive power
  of the median v3 models for the magnification of sources vs. that
  seen in v4 parametric models.  The shaded red region shows $1\sigma$
  scatter in the v4 magnifications in the median v4 magnification
  vs. the magnification factors reported in various literature
  studies.  Below a model magnification factor of $\sim$50, both
  results in the literature and the median v3 models agree very well
  with the v4 model results; above a magnification factor of 50, the
  magnification factor for sources, as estimated by the v4 models,
  saturates around values of $\sim$50 to 100.  Both the points and
  shaded regions for magnification factors $>$50 are shown with a
  lighter shading to reflect the lesser reliability of magnification
  factors in this regime.  (\textit{right}) Similar to left figure,
  but showing the ratio of the median v4 magnification and the
  reported magnification factors.\label{fig:predictlit}}
\end{figure*}

\begin{figure*}
\epsscale{1.17}
\plotone{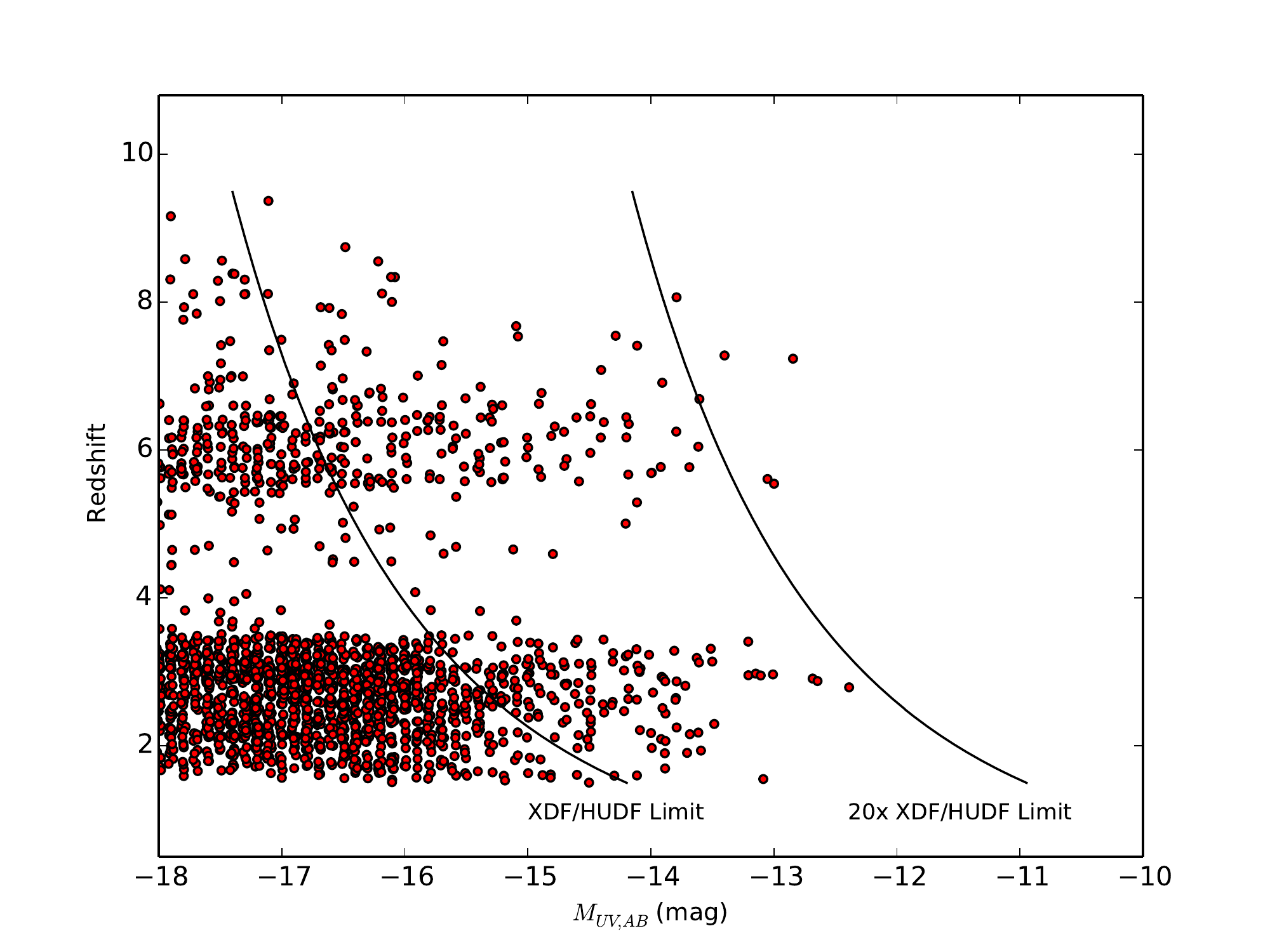}
\caption{Estimated $UV$ luminosity of candidate star-forming galaxies
  in our HFF samples vs. redshift (\textit{filled red circles}).  The
  two solid lines show the $UV$ luminosities probed by the 30-mag
  XDF/HUDF data and $20\times$ fainter than the XDF/HUDF data.
  Clearly, the full HFF data set contains a large numbers of candidate
  ultra-faint ($\gtrsim -$14 mag) galaxies from $z\sim8$ to $z\sim2$.
  The much smaller number of sources in our samples at $z\sim3.5$-5.5
  is a direct result of the much more conservative criteria required
  to select such sources free of contamination.\label{fig:muv_vs_red}}
\end{figure*}

In general, the model magnification factors predict the ``true''
magnification factors to a magnification factor of $\sim$20-30 and
then lose their predictive power at magnification factors of $\sim$20
(MACS1149), $\sim$30 (MACS0717) to $\sim$50-100 (Abell 2744, MACS
0416, Abell 370, Abell S1063).  The predictive power of the individual
v4 models appears to be higher, as expected, than the v1 models for
Abell 2744, MACS 0416, Abell 370, and Abell S1063, with the most
significant improvement seen for the Abell S1063 models.

\begin{deluxetable*}{ccccccccccc}
\tablewidth{0cm}
\tablecolumns{11}
\tabletypesize{\footnotesize}
\tablecaption{Samples of $z=2$-10 Galaxies found over the six HFF cluster fields (including Oesch et al.\ 2018a selection)\tablenotemark{b}\label{tab:samp}}
\tablehead{\colhead{Cluster} & \colhead{Area [arcmin$^2$]} & \colhead{$z\sim 2$} & \colhead{$z\sim 3$} & \colhead{$z\sim 4$} & \colhead{$z\sim 5$} & \colhead{$z\sim 6$} & \colhead{$z\sim 7$} & \colhead{$z\sim 8$} & \colhead{$z\sim 9$} & \colhead{$z\sim 10$\tablenotemark{b}}}
\startdata
Abell 2744 & 4.9  & 157 & 233 & ---\tablenotemark{a} & 27\tablenotemark{c} & 49 & 25 & 15 & 4 & 2\tablenotemark{b}\\
MACS0416 & 4.9  & 215 & 233 & ---\tablenotemark{a} & 7\tablenotemark{c} & 50 & 26 & 10 & 6 & 0\\
MACS0717 & 4.9  & 81 & 160 & 32\tablenotemark{c} & ---\tablenotemark{a} & 26 & 14 & 9 & 0 & 0\\
MACS1149 & 4.9  & 134 & 195 & 36\tablenotemark{c} & ---\tablenotemark{a} & 52 & 21 & 5 & 2 & 0\\
Abell S1063 & 4.9  & 96 & 203 & ---\tablenotemark{a} & 11\tablenotemark{c} & 62 & 28 & 6 & 3 & 0\\
Abell 370 & 4.9  & 82 & 152 & ---\tablenotemark{a} & 14\tablenotemark{c} & 35 & 11 & 6 & 1 & 0\\
Total & 29.4 & 765 & 1176 & 68 & 59 & 274 & 125 & 51 & 16 & 2
\enddata
\tablenotetext{a}{Sources are not selected at this redshift in the indicated cluster field, due to concerns about contamination from foreground galaxies from the cluster due to the similar position for the spectral break (Figure~\ref{fig:clust_contam}).}
\tablenotetext{b}{From Oesch et al.\ (2018a).  See also Zitrin et al.\ (2014).}
\tablenotetext{c}{$z\sim4$ and $z\sim5$ star-forming galaxies only selected brightward of 27.3 and 27.5 mag, respectively, to minimize the impact of contamination from foreground cluster galaxies on our results (see \S\ref{sec:z45} and Figure~\ref{fig:purity}).}
\end{deluxetable*}

In Figure~\ref{fig:predictind}, we also show how well the median of
the v4 magnification models predict the magnification factors of the
v4 magnification model not included in the median.  For all six HFF
clusters, the median magnification model is successful at predicting
the ``true'' magnification to values of $\gtrsim$40, with the best
performance achieved for Abell S1063, with the predictive power
extending to values of $\sim$100.

In Figure~\ref{fig:predict}, we show how well the median of the v1,
v3, and v4 magnification models predict the magnification factors of
the v4 models.  For five of the six clusters, except perhaps Abell
2744, there is a clear improvement in the predictive power of the
median v4 model over the median v1 model.  For four of the clusters,
i.e., Abell 2744, MACS 0416, Abell 370, and Abell S1063, the median of
the v4 magnification models is successful at predicting the
magnification factors to values of $\gtrsim$80.

A number of previous studies have explicitly tabulated the
magnification factors they derived for $z\sim6$-10 galaxies (Zitrin et
al.\ 2014; Infante et al.\ 2015; Livermore et al.\ 2017; Ishigaki et
al.\ 2018) using either their own lensing models or the publicly
available HFF lensing models.  As all but one of these models predated
the latest v4 models and made use of a smaller set of constraints,
comparison of these published magnification factors with the v4
results provide us with a measure of the robustness of the
magnification factors, after improvements are made to the models.

Figure~\ref{fig:predictlit} compares various published magnification
factors from the literature (Zitrin et al.\ 2014; Infante et
al.\ 2015; Livermore et al.\ 2017; Ishigaki et al.\ 2018) with the
median of the v4 parametric models.  The green-shaded region shows the
expected $1\sigma$ range in predictive power of the median v3 models
for the magnification of sources vs. that seen in v4 parametric
models.  The green-shaded region is a marginalized version of the
green shaded region shown in Figure~\ref{fig:predictind} over the
first four clusters (which feature v3 models).  The red shaded area
shows the $\pm$1$\sigma$ range in median v4 magnification factors
found for sources within a 0.2-dex range of published magnification
factors.

Encouragingly, we find that the published magnification factors are
consistent with the median v4 magnifications to values of $\sim$40-50.
For higher magnification factors, i.e., $>$60, we find that the
reported magnifications magnifications are less robust, and the median
magnifications we find for those sources using the v4 parametric
models lie in the range $\sim$40 to $\sim$100.

\subsubsection{The Challenge of Identifying Especially Faint Galaxies}

One area where both new models make quite a difference is in our quest
for the faintest galaxies.  These are important for helping constrain
both the faint end of the luminosity function and also for the role
they play in distinguishing various galaxy formation scenarios.  The
challenge, however, of working at high magnifications in a regime
where the models are both improving and evolving can be seen from the
results for the very faintest objects.  As such, it is interesting to
reconsider the faintest sources found with earlier models to have
exceptionally low luminosities, i.e., $\gtrsim-15$ at $z\sim6$-9 and
$\gtrsim-16$ at $z\sim9$.  There are $>$20 such sources presented thus
far in the literature.  For this exercise, we use the same methods and
lensing models as we use here.

\begin{figure*}
\epsscale{1.07}
\plotone{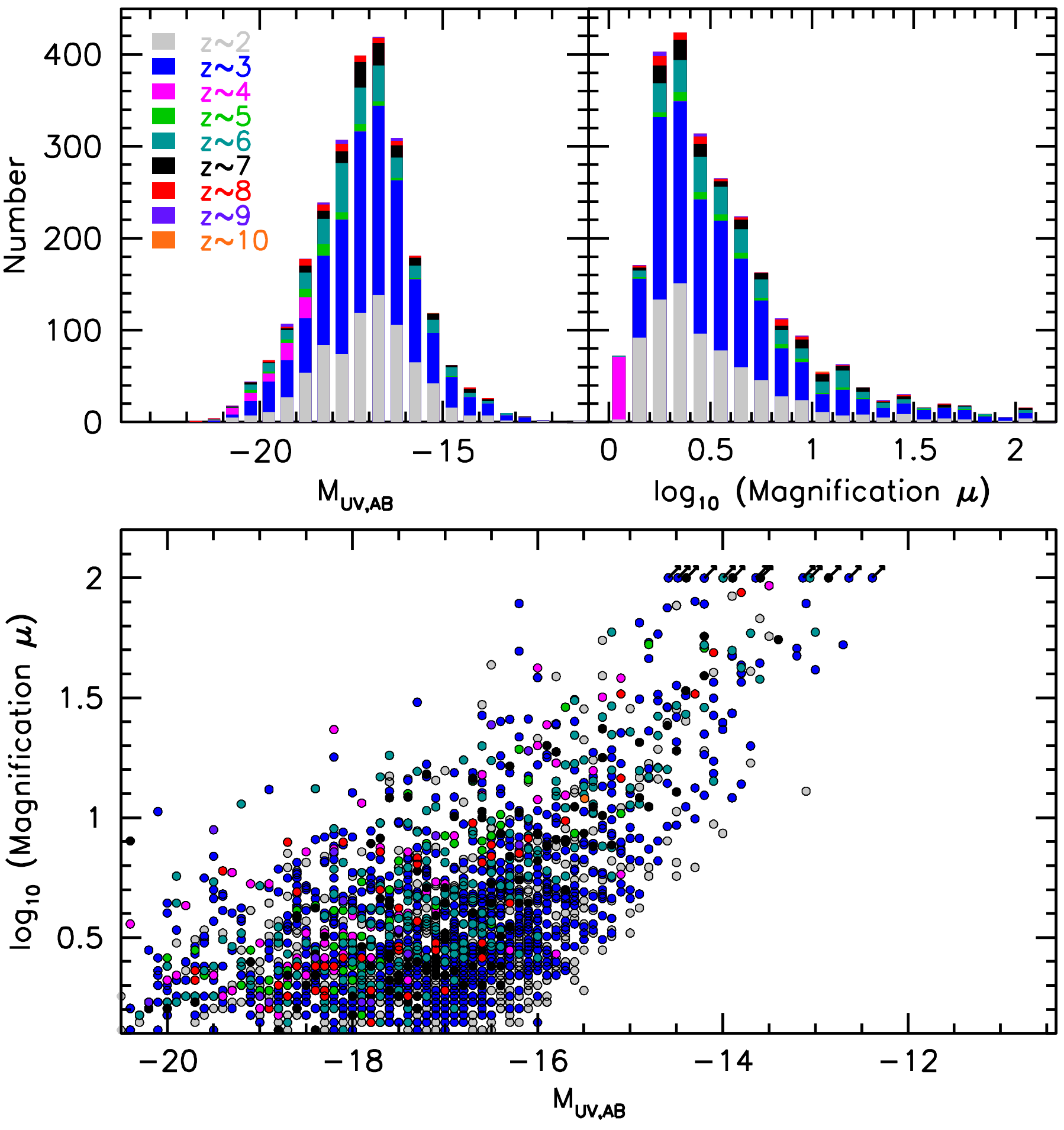}
\caption{Number of $z\sim2$, $z\sim3$, $z\sim4$, $z\sim5$, $z\sim6$,
  $z\sim7$, $z\sim8$, $z\sim9$, and $z\sim10$ sources (\textit{grey,
    blue, magenta, green, cyan, black, red, purple, and orange
    histograms, respectively}) identified over all six HFF clusters
  and parallel fields vs. the inferred absolute magnitude
  (\textit{upper left panel}) and model magnification factor
  (\textit{upper right panel}).  We impose a hard upper limit of 100
  on the allowed magnification factor of sources, affecting just 16
  sources ($<$1\% of the total).  (\textit{lower panel}) Median model
  magnification factors vs. absolute magnitudes for sources in our
  $z\sim2$, $z\sim3$, $z\sim4$, $z\sim5$, $z\sim6$, $z\sim7$,
  $z\sim8$, $z\sim9$, and $z\sim10$ samples (\textit{grey, blue,
    magenta, green, cyan, black, red, purple, and orange points,
    respectively}).\label{fig:hist}}
\end{figure*}

What we find is that the faintest galaxies from several previous
studies, those reported to be fainter than approximately $-$15.5 mag
and $-$14.5 mag, are $\sim$0.5 mag and 0.7 mag brighter, respectively.
In the median, the faint sources are estimated to be 0.4 mag and 0.5
mag brighter, respectively.  In addition, we find that few of the
sources with nominal $UV$ luminosities fainter than $-$14 mag are
estimated to be fainter than $-$14 mag using the median v4 parametric
models.

Even with the v4 updates to the magnification estimates to individual
sources, the latest models will likely be revised in the future.  As
such, caution is required in considering the implications of the
faintest sources in the current compilation.  While use of median
magnification models should improve the robustness of the
magnification estimates, many of the models make use of similar
assumptions and similar observational constraints. As a result, the
true systematic uncertainties could well be larger than indicated by
the dispersion in model results.

\subsubsection{Fiducial Magnification Factors}

For our fiducial estimate of the magnification for individual sources,
we take the median of the magnification estimates from the v4
parametric models, i.e., \textsc{GLAFIC}, \textsc{CATS},
Sharon/Johnson, Zitrin-NFW, Keeton, and Caminha (where available)
since the parametric models in general have proven to be among the
best performing models for the HFF comparison project (Meneghetti et
al.\ 2017).  Included in our estimates of the median magnification
factor are the best $\kappa$ and $\gamma$ maps and the other models in
the associated Markov Chain Monte Carlo (MCMC) chain.  Based on the
range in magnifications from the parametric models, we estimate 68\%
confidence intervals on the model magnification factors.  In cases
where both a v4 and v4.1 model exists, both are considered equally in
computing a median for that specific flavor of lensing model.

While the tests in the previous subsection support the robustness of
magnification estimates to factors of $\gtrsim$40 and have utility in
predicting the magnification factors to values of $\sim$100, they also
cast doubt on the robustness of these estimates when the magnification
factors exceed 100.  In the interest of maximizing the robustness of
the results we present, we adopt a maximum magnification factor of 100
for our analyses and also for our companion paper on the $UV$ LFs
(R. Bouwens et al.\ 2022, in prep).  For the few sources in our
catalogs where the median magnification factors we derive exceed 100,
i.e., 16 sources, we explicitly set the magnification estimates of
those sources equal to 100 (as can also be seen in
Figure~\ref{fig:hist} from \S\ref{sec:finalsamples}).

\begin{deluxetable*}{cccccccc}
\tablewidth{0pt} \tablecolumns{8} \tabletypesize{\footnotesize}
\tablecaption{A complete list of the sources included in the $z\sim2$,
  $z\sim3$, $z\sim 4$, $z\sim 5$, $z\sim 6$, $z\sim 7$, $z\sim 8$,
  $z\sim9$, and $z\sim 10$ samples from the present selection and that
  of Oesch et al.\ 2018a\tablenotemark{*}\label{tab:catalog}}
\tablehead{ \colhead{ID} & \colhead{R.A.} & \colhead{Dec} &
  \colhead{$m_{AB}$\tablenotemark{a}} & \colhead{Sample\tablenotemark{b}} &
  \colhead{Data Set\tablenotemark{c}} &
  \colhead{$z_{phot}$\tablenotemark{d}} & \colhead{$\mu_{Median-Parametric}$}\tablenotemark{e}} \startdata 
 A2744275-4227525020 & 00:14:22.75 & $-$30:25:02.0 & 27.45 &  2 & 17 &  2.44 &   2.6$_{-0.3}^{+0.1}$\\
 A2744275-4236124532 & 00:14:23.61 & $-$30:24:53.2 & 28.45 &  2 & 17 &  2.33 &   3.2$_{-0.3}^{+0.3}$\\
 A2744275-4245824489 & 00:14:24.58 & $-$30:24:48.9 & 28.22 &  2 & 17 &  2.30 &   3.0$_{-0.1}^{+0.4}$\\
 A2744275-4239224503 & 00:14:23.92 & $-$30:24:50.3 & 25.28 &  2 & 17 &  1.93 &   3.3$_{-0.2}^{+0.4}$\\
 A2744275-4240424495 & 00:14:24.04 & $-$30:24:49.5 & 24.92 &  2 & 17 &  1.74 &   3.2$_{-0.2}^{+0.4}$\\
 A2744275-4224124478 & 00:14:22.41 & $-$30:24:47.8 & 28.60 &  2 & 17 &  2.30 &   4.6$_{-0.6}^{+0.3}$\\
 A2744275-4239924345 & 00:14:23.99 & $-$30:24:34.5 & 28.00 &  2 & 17 &  2.15 &   6.7$_{-1.4}^{+1.6}$\\
 A2744275-4225724288 & 00:14:22.57 & $-$30:24:28.8 & 26.26 &  2 & 17 &  2.10 &   3.7$_{-0.3}^{+0.3}$\\
 A2744275-4230424248 & 00:14:23.04 & $-$30:24:24.8 & 29.49 &  2 & 17 &  2.21 &   4.8$_{-0.3}^{+0.8}$\\
 A2744275-4226724234 & 00:14:22.67 & $-$30:24:23.4 & 28.91 &  2 & 17 &  1.95 &   2.5$_{-0.2}^{+0.7}$
\enddata
\tablenotetext{*}{Table~\ref{tab:catalog} is published in its entirety
  in the electronic edition of the Astrophysical Journal.  A portion
  is shown here for guidance regarding its form and content.}
\tablenotetext{a}{The quoted apparent magnitude includes a correction for the model magnification factor.}
\tablenotetext{b}{The mean redshift of the sample in which the source
  is included.}
\tablenotetext{c}{The data set from which the source was selected: 
  17 = Abell2744, 18 = MACS0416, 19 = MACS0717, 20 = MACS1149, 21 = Abell S1063, 
and 22 = Abell 370} 
\tablenotetext{d}{Most likely redshift in
  the range $z=1.5$-11 as derived using the EAzY photometric redshift
  code (Brammer et al.\ 2008) using the same templates as discussed in
  \S\ref{sec:selection}.}  
\tablenotetext{e}{Median of the magnification factor computed using the latest parametric HFF models (see Table~\ref{tab:models}).}
\end{deluxetable*}

\subsection{Final Samples of $z\sim2$-9 Galaxies\label{sec:finalsamples}}

The present procedure resulted in the construction of very large
samples of $z\sim2$-9 galaxies.  Our $z\sim2$, $z\sim3$, $z\sim4$,
$z\sim5$, $z\sim6$, $z\sim7$, $z\sim8$, and $z\sim9$ samples include
765, 1176, 68, 59, 274, 125, 51, and 16 sources, respectively.  The
total number of sources in the collective $z\sim2$-10 sample from this
work and that of Oesch et al.\ (2018a) is 2536.  These results are
summarized in Table~\ref{tab:samp}.

With our total sample of sources stretching from $z=2$ to $z=10$, it
is interesting to show the luminosities that we potentially probe with
HFF program.  In Figure~\ref{fig:muv_vs_red}, we present the $UV$
luminosities of sources in our selection vs. the photometric redshift
we derive from EAzY.  Remarkably, for our selection (and that of Oesch
et al.\ 2018a), we probe to $-$12.4 mag at $z\sim3$ and $-$12.9 mag at
$z\sim7$.  For context, we also present a black line illustrating the
$UV$ luminosity that we probe at 30 mag in the eXtreme Deep Field
(XDF) / {\it Hubble} Ultra Deep Field (HUDF: Beckwith et al.\ 2006;
Illingworth et al.\ 2013; Koekemoer et al.\ 2013).  From this figure,
it is clear that with the HFFs, we probe $\gtrsim$20$\times$ fainter
than what is possible in the XDF/HUDF.

We give further insights into our sample in Figure~\ref{fig:hist}.
The upper left and right panels show the cumulative number of
star-forming galaxies at $z=2$, 3, 4, 5, 6, 7, 8, 9, and 10 as a
function of $UV$ luminosity and magnification, respectively, while the
distribution in magnification and $UV$ luminosity is shown in lower
panel of the same figure.  A complete compilation of the present
sample of $z=2$-9 galaxy candidates is provided in
Table~\ref{tab:catalog}.

The present selection of $z\sim2$-9 galaxies is $\sim$2-3$\times$
larger than previous selections at these redshifts.  Previously, Alavi
et al.\ (2017) had identified 297 z$\sim$1.0-1.6, 318 z$\sim$1.6-2.2,
and 278 z$\sim$2.2-3.0 galaxies using the {\it HST} observations over
three lensing clusters: Abell 1689 and two HFF clusters Abell 2744 and
MACS0717.  At $z\geq 6$, Ishigaki et al.\ (2018) had identified 140
$z\sim6$-7, 27 $z\sim8$, and 14 $z\sim9$ galaxies, respectively, over
all six clusters.  Atek et al.\ (2018) report finding 300 $z\sim6$-7
galaxies over the six HFF clusters, while Leung et al.\ (2018) report
finding 260 $z\geq 4.75$ galaxies.  Yue et al.\ (2018) report a total
sample of 272 $z=5$-9.5 galaxies from the first four HFF clusters.
Atek et al.\ (2015b) had identified some 119 $z\sim6$-7 galaxies and 13
$z\sim8$ galaxies as part of their HFF selections over three clusters,
while Livermore et al.\ (2017) identified 105 $z\sim6$, 40 $z\sim7$,
and 16 $z\sim8$ galaxies behind Abell 2744 and MACS0416.  Zheng et
al.\ (2014), Infante et al.\ (2015), Laporte et al.\ (2016), Zheng et
al.\ (2017) report 117 $z=6$-10 sources in total from the first 4 HFF
cluster and parallel fields.

\section{$z=2$-9 Galaxy Surface Density Results}

The purpose of this section will be to quantify the dependence of the
surface density of lensed $z=2$-9 galaxies on the magnification factor
and then to use the results to draw conclusions about the physical
characteristics of faint $z=2$-9 galaxies.  The dependence of the
surface density of lensed galaxies on the magnification factor is
frequently referred to as the magnification bias in the literature
(Turner et al.\ 1984; Broadhurst 1995, Broadhurst et al.\ 2005).

The magnification bias is often expressed as the ratio $N'(<m)/N_o
(<m) = \mu^{2.5 S - 1}$ where $N'$ and $N_o$ represent the surface
density of galaxies to some fixed apparent magnitude limit $m$ in
unlensed and lensed observations respectively, $\mu$ is the
magnification factor, and $S$ is the intrinsic count slope $d\log
N(<m)/dm$ (Broadhurst 1995).  The $\mu^{-1}$ in the above equation
accounts for the reduction in volume available to find sources in a
given magnified volume of the universe.

For sources at a fixed redshift and amplified by a magnification
factor $\mu$, $N(<m)$ is the integral of the LF to some magnitude
limit $m$ of a survey $\int _{M>M_{lim},z} \phi(M) P(m,z,\mu)
(dV/d\Omega) dz dM$ where $M$ is the absolute magnitude of sources,
$P(m,z,\mu)$ is the selection efficiency as a function of apparent
magnitude $m$, redshift $z$, and magnification factor $\mu$, and
$dV/d\Omega$ is the comoving volume per solid angle $\Omega$.  Making
the simplifying assumption that each of the factors in the integral
are independent, the intrinsic slope $S$ is then $d \log
\int_{M<M_{lim}} \phi(M) dM$.  For a power law LF $\phi(M) \propto
10^{-0.4 (\alpha+1) (M-M^*)}$, this reduces to $-0.4(\alpha+1)$, such
that $N'(<m)/N_o (<m) = \mu^{-2-\alpha}$.  In the more general case
that the source selection efficiency $S(\mu)$ depends on the
magnification factor, the relevant expression is $\mu^{-2-\alpha}
S(\mu)$.

We will start off this section by expanding on the simple discussion
just provided in the previous two paragraphs, describing several
significant effects which impact the measured surface densities seen
behind lensing clusters and drive a dependence on the magnification.
We will then quantify how the surface densities of $z=2$-7 galaxies in
our selected samples depend on the model magnification factors.

\begin{figure*}
\epsscale{1.17}
\plotone{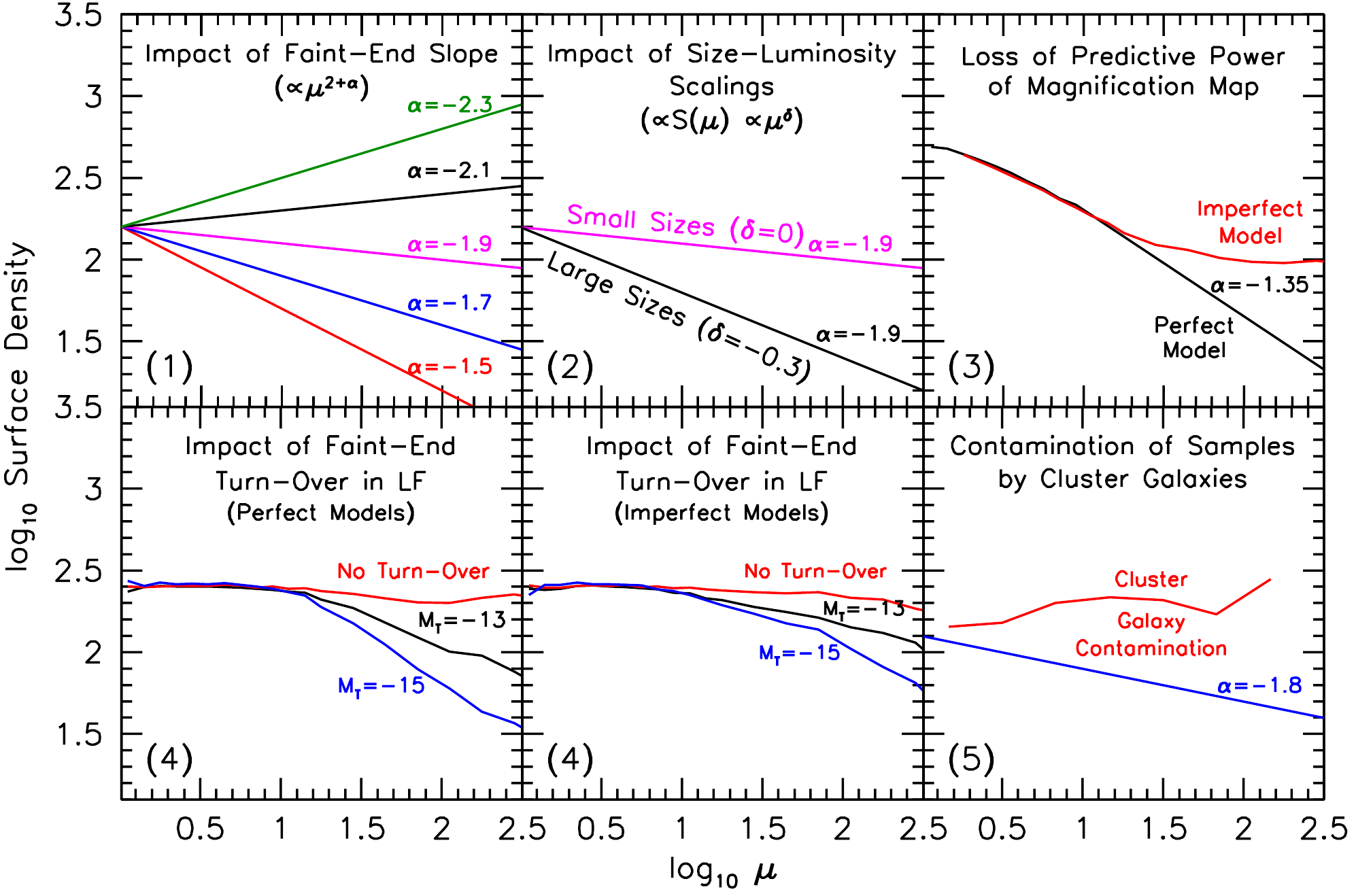}
\caption{Schematic illustrating how the surface density $\Sigma$ of
  galaxies would be expected to depend on the magnification factor
  $\mu$, as outlined in the five subsections of \S4.1: the relevant
  subsection number is indicated in each panel.  Five of the six
  panels illustrate a different issue that impacts the dependence.  As
  shown in \S3 of Bouwens et al.\ (2017: see also Broadhurst 1995,
  Broadhurst et al.\ 2005), the observed surface densities is expected
  to vary as $\mu^{2+\alpha} S(\mu)$ where $S(\mu)$ is the
  completeness of faint selections as a function of magnifications
  $\mu$.  The numbering here and in the figure corresponds to the
  numbering in \S4.1.  (1) The upper left panel shows the expected
  dependence on the faint-end slope $\alpha$ of the LF (with assumed
  $\alpha$'s indicated on the figure), while (2) the upper middle
  panel shows the expected dependence on different size-luminosity
  relations, while keeping the assumed faint-end slope $\alpha$ fixed
  to $-$1.9.  Different size-luminosity relations translate into
  different magnification-dependent selection efficiencies $S(\mu)$
  which we model as $\mu^{\delta}$.  The large size-luminosity
  relation shown here relies on the $S(\mu)\propto\mu^{-0.3}$
  dependence found by Oesch et al.\ (2015), i.e., equivalent to
  $\delta = -0.3$ in the chosen parameterization, while the small
  size-luminosity relation assumes that sources are unresolved (e.g.,
  Bouwens et al.\ 2017a; Kawamata et al.\ 2018; Bouwens et al. 2021b,
  2022), resulting in a magnification-independent selection
  efficiency, i.e., $\delta=0$. (3) The upper right panel shows the
  expected flattening to the $\Sigma$ vs. model $\mu$ relationship,
  due to a breakdown in the predictive power of the models (see Figure
  5 of Bouwens et al.\ 2017b where this result was previously shown)
  and assumes a faint-end slope of $-1.35$ to the $UV$ LF.  (4) The
  lower left and center panels presents the impact of a turn-over in
  the $UV$ LF at the faint end, with no faint-end turn-over
  (\textit{red}), a turn-over $M_T$ at $-15$ mag (\textit{black}), and
  a turn-over $M_T$ at $-13$ mag (\textit{blue}), respectively.  The
  lower left panel shows the dependence assuming perfect recovery of
  the magnification factor, while the lower center panel shows the
  dependence in a more realistic scenario where the magnifications are
  uncertain.  (5) The lower right panel contrasts the expected
  dependence for $z\sim5$ galaxies (assuming a faint-end slope
  $\alpha$ of $-1.8$) with that expected including large numbers of
  foreground galaxies from the clusters Abell 2744 and MACS0416.  The
  selection efficiency is assumed to be independent of magnification,
  i.e., $\delta=0$, except in the upper center
  panel.\label{fig:surfddd}}
\end{figure*}

\subsection{Expected Dependencies on the Source Magnification Factor}

There are at least five distinct effects which can impact the observed
surface density of galaxies vs. magnification factor $\mu$.  Each of
these effects is illustrated in Figure~\ref{fig:surfddd} and discussed
in the paragraphs that follow:\\

\noindent \textit{(1) Dependence on the Faint-end Slope $\alpha$:} As
we demonstrated earlier, it is well established (e.g. Broadhurst 1995)
that the surface density of sources in our high-redshift samples scale
as $\mu^{-(2+\alpha)}$ (e.g., Broadhurst 1995) assuming a Schechter
  form for the $UV$ LF.  For this first case, we assume that source
  detectability does not depend on the magnification factor.  Also
  assumed is that the model magnification factor is exactly equal to
  the true magnification factor everywhere.

\begin{figure*}
\epsscale{1.17}
\plotone{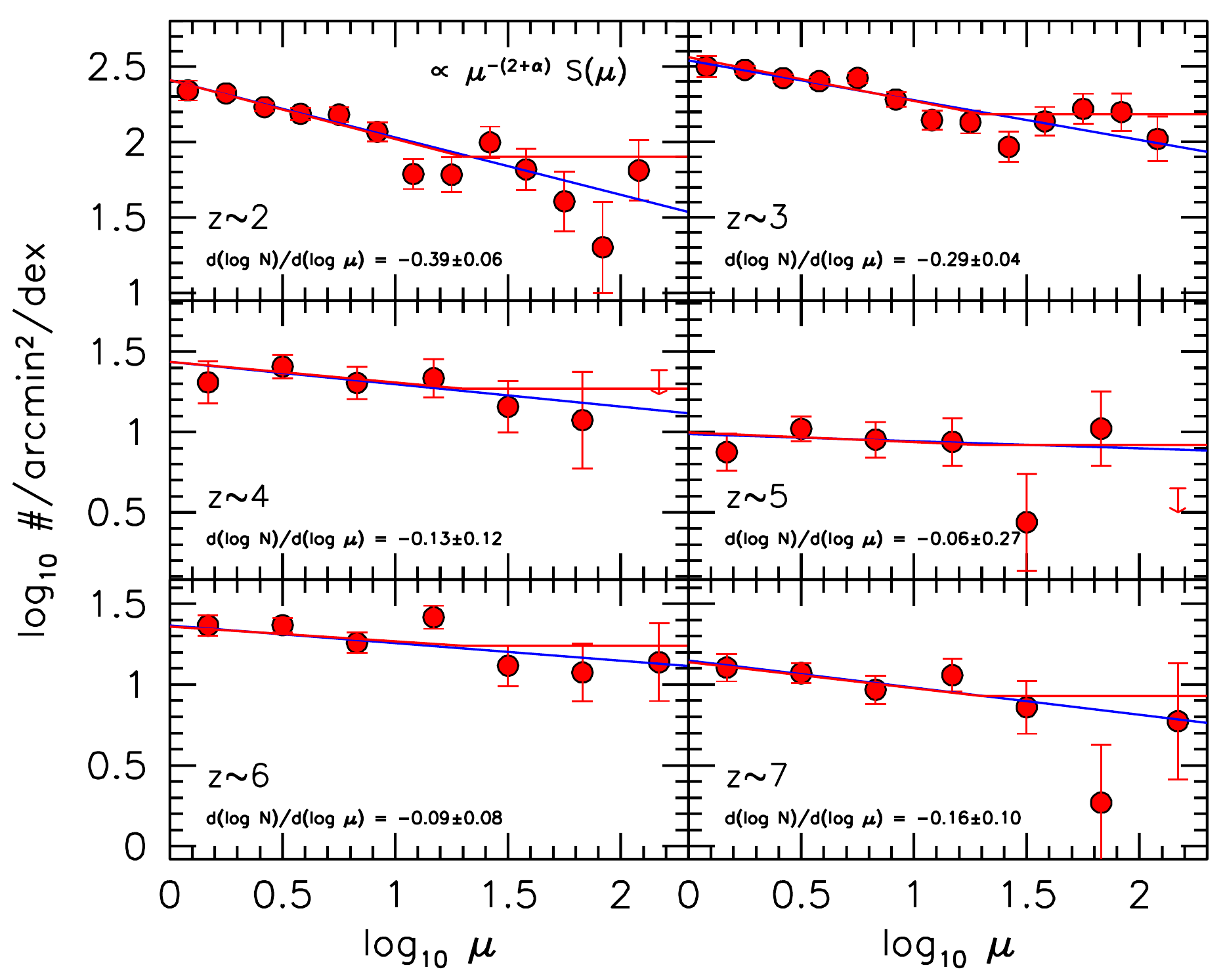}
\caption{The surface density of star-forming galaxies in our $z\sim2$,
  $z\sim3$, $z\sim4$, $z\sim5$, $z\sim6$ and $z\sim7$ selections
  (\textit{each shown in their own panel}) vs. the median
  magnification factor from the parametric lensing models.  The
    surface density is plotted per unit magnification factor (dex).
  The blue and red lines represent power-law fits to the binned red
  points, assuming a break-down in the predictive power of the lensing
  models at values greater than 300 and 30, respectively, i.e.,
  $\mu_{break} = 300$ and 30.  As shown in \S3 of Bouwens et
  al.\ (2017b), we would expect the observed surface densities to vary
  as $\mu^{-(2+\alpha)} S(\mu)$ where $S(\mu)$ is the completeness of
  faint selections as a function of magnifications $\mu$.  As
  discussed in \S4.2 and \S4.3 and as shown in
  Figure~\ref{fig:faintend}, we see that such expectations are, in
  fact, realized, and there is a connection between the slopes of the
  lines plotted here and the best estimate faint-end slopes $\alpha$
  observed in field studies.\label{fig:surfd}}
\end{figure*}

For faint-end slopes $\alpha$ steeper than $-2$, the surface density
of sources should be the highest in the highest magnification regions.
For faint-end slopes of $-2$, the surface density of sources is
independent of the magnification factor $\mu$.  For faint-end slopes
shallower than $-2$, the surface density of sources is highest in the
lowest magnification regions.  The upper-left panel of
Figure~\ref{fig:surfddd} illustrates this expected dependencies, with
faint-end slopes $\alpha$ of $-1.5$ (\textit{red line}), $-1.7$
(\textit{blue line}), $-1.9$ (\textit{magenta line}), $-2.1$
(\textit{black line}), and $-2.3$ (\textit{green line}).  The
sensitivity of the results to the faint-end slope $\alpha$ is quite
strong.\\

\noindent \textit{(2) Sizes and Surface Brightnesses of
  Lower-Luminosity Galaxies:} In practice, the surface density we
measure of lensed galaxies behind a cluster depends on their detection
efficiency, which in turn can depend on the sizes or surface
brightnesses of galaxies.  For extended sources, Oesch et al.\ (2015)
demonstrated that the detection efficiency $S(\mu)$ could show a
noteworthy dependence on the magnification factor $\mu$, finding
approximately an $\mu^{-0.3}$ dependence (e.g., as seen in Figure 3 of
Oesch et al.\ 2015) for a model where the sizes are proportional to
the luminosity $L$ to the 0.22 power (i.e., $L^{0.22}$).  Bouwens et
al.\ (2017a), Bouwens et al.\ (2017c), Kawamata et al.\ (2018),
Bouwens et al.\ (2021b), and Bouwens et al.\ (2022) have argued that
high-redshift sources at very low luminosities are very compact.  If
this is in fact the case, we would expect the detection efficiency of
sources to show only a marginal dependence on the magnification factor
$\mu$ and to depend only on the redshift and apparent magnitude of
sources.

Incorporating the impact of the selection efficiency, we expect the
surface densities to depend on the magnification factor $\mu$ as
$\mu^{-(2+\alpha)} S(\mu)$.  The upper-middle panel of
Figure~\ref{fig:surfddd} illustrates the expected dependence of the
surface densities on the magnification factor $\mu$ for especially
compact sources where $S(\mu) \propto \mu^0$ (\textit{magenta line})
and extended sources, such as Oesch et al.\ (2015) assumed in their
simulations, where $S(\mu) \propto \mu^{-0.3}$ (\textit{black line}).
A faint-end slope $\alpha$ of $-1.9$ is assumed in both cases.\\

\noindent \textit{(3) Breakdown in Predictive Power of Lensing Models
  at high magnification factors $\mu$:} Another important effect
regards differences between the ``true'' magnification factors and
model magnification factors.  Given that the observed surface
densities depend on the ``true'' magnification factors and not the
model magnification factors, any breakdown in the relationship between
the true and model magnification factors would impact the dependence
the surface densities show on the estimated magnification factors.

Such a breakdown is expected to occur at magnification factors in
excess of 50 to 100, as we for example show in
Figures~\ref{fig:predictind}-\ref{fig:predictlit} (see also Meneghetti
et al.\ 2017; Bouwens et al.\ 2017b).  This effect would cause any
dependence of surface density on the model magnification factor to be
effectively washed out, resulting in galaxy surface densities
asymptoting to a fixed value.

We illustrate such a breakdown in the dependence of the measured
surface densities on the model magnification factor with the red line
in the upper-right panel of Figure~\ref{fig:surfddd}.  If no such
breakdown occurs and the model magnification maps can be effectively
used to magnification factors of 300, the dependence would look like
the black line in that panel.  For this particular example, we adopt
the results shown in the center panel of Figure 5 of Bouwens et
al.\ (2017b), where the underlying model LF is assumed to have a
faint-end slope $\alpha$ of $-1.35$, the observations are set up using
the \textsc{GLAFIC} magnification models, and the recovered
magnifications are from the CATS magnification models.

One consequence of this is for the recovered $UV$ LF to asymptote to a
faint-end slope of $-2$ for a magnification-independent selection
efficiency and $-2+d(\ln S(\mu))/d(\ln \mu)$ in the more general case
(Appendix C of Bouwens et al.\ 2017b).\\

\noindent \textit{(4) Faint-end Turn-over in the $UV$ LF:} A turnover
in the UV LF at the faint end also impacts the dependence of source
surface density on the magnification factor.  Such a turn-over
primarily impacts the number of sources found in the highest
magnification regions probing the faintest sources and thus shows up
as a decrease in the surface density of sources at high magnification
factors.

As an illustration of the dependence expected, the predicted
dependence on magnification factor is shown in
Figure~\ref{fig:surfddd} (\textit{lower left and lower center panels})
for $UV$ LFs with no faint-end turn-over and turn-overs at $-15$ mag
and $-13$ mag.  We implemented these turn-overs using the functional
form earlier presented in Bouwens et al.\ (2017b), assuming the CATS
v4.1 magnification model, and assuming perfect recovery of the
magnification factors from the model in one case (\textit{lower left
  panel}) and a more realistic recovery in a second case
(\textit{lower center panel}) using a median of the parametric models
not including CATS.  As is clear from the panels, the LFs with a
faint-end turn-over show a break in the surface densities toward lower
values at magnification factors higher than 15.  In the cases where
the magnification is uncertain (\textit{lower center panel}), the
influence of the turn-over on the surface density of galaxies at high
magnifications is less obvious.  Interested readers may also want to
consult an earlier discussion on this topic by Leung et al.\ (2018).\\

\noindent \textit{(5) Contamination of $z\gtrsim2$ Selections with
  Cluster Galaxies:} The measured surface densities of $z\gtrsim2$
selections vs. magnification factor can also be impacted if foreground
galaxies significantly contaminate selections.  While essentially all
selections suffer from low levels of contamination, contamination can
become much more serious when search fields have large numbers of
foreground cluster galaxies that show Balmer or 4000\AA$\,$breaks
across exactly the same passbands as the Lyman-break in a given search
(e.g., see Figure~\ref{fig:clust_contam}).  For the HFF clusters, this
can be a serious concern for $z\sim4$ and $z\sim5$ selections
depending on the cluster redshift (see Figure~\ref{fig:clust_contam}).

\begin{deluxetable}{ccc}
\tabletypesize{\footnotesize}
\tablecaption{Dependence of the Source Surface Density on the Magnification Factors and the Direct Implications\label{tab:slope}}
\tablehead{
\colhead{Redshift} & \colhead{Measured} & \colhead{Inferred} \\
\colhead{Sample} & \colhead{$\epsilon$\tablenotemark{a}} & \colhead{$\alpha$\tablenotemark{b}}}
\startdata
\multicolumn{3}{c}{Assuming Lensing Model Predictive Power $\mu_{break} = 300$, $\delta=0$}\\
z$\sim2$ & $-$0.38$\pm$0.05 & $-$1.62$\pm$0.05\\
z$\sim3$ & $-$0.26$\pm$0.04 & $-$1.74$\pm$0.04\\
z$\sim4$ & $-$0.14$\pm$0.14 & $-$1.86$\pm$0.14\\
z$\sim5$ & $-$0.05$\pm$0.14 & $-$1.95$\pm$0.14\\
z$\sim6$ & $-$0.12$\pm$0.08 & $-$1.88$\pm$0.08\\
z$\sim7$ & $-$0.17$\pm$0.11 & $-$1.83$\pm$0.11\\
\\
\multicolumn{3}{c}{Assuming Lensing Model Predictive Power $\mu_{break} = 30$, $\delta=0$}\\
z$\sim2$ & $-$0.39$\pm$0.06 & $-$1.61$\pm$0.06\\
z$\sim3$ & $-$0.29$\pm$0.04 & $-$1.71$\pm$0.04\\
z$\sim4$ & $-$0.13$\pm$0.15 & $-$1.87$\pm$0.15\\
z$\sim5$ & $-$0.06$\pm$0.16 & $-$1.94$\pm$0.16\\
z$\sim6$ & $-$0.10$\pm$0.09 & $-$1.90$\pm$0.09\\
z$\sim7$ & $-$0.16$\pm$0.12 & $-$1.84$\pm$0.12
\enddata
\tablenotetext{a}{Slope of the best-fit surface density vs. magnification factor $\mu$ relations (Figure~\ref{fig:surfd}).}
\tablenotetext{b}{Computed assuming $\delta=0$.  If the selection
  efficiencies were lower at high magnification factors (as one would
  expect given standard size-luminosity relations: e.g., Shibuya et
  al.\ 2015), i.e., $\delta=-0.3$, the faint-end slopes we derive would
  be even steeper than the values provided in this table.  For
  example, utilizing the Oesch et al.\ (2015) selection efficiency
  vs. magnification factor $\mu$ scaling, we would infer faint-end
  slopes that were $\Delta\alpha$ $\sim$0.3 steeper.}
\end{deluxetable}

Any significant contamination of our high-redshift samples by cluster
galaxies would impact the surface density vs. magnification $\mu$
relation.  The surface density of cluster galaxies is much higher
towards the center of a cluster where the magnifications are higher
than it is towards the outer parts of a cluster.  If cluster galaxies
substantially contaminated a selection, one would expect the surface
density of galaxies to rise appeciably to magnification factors of 10
and perhaps flatten beyond that.

To provide an approximate illustration of what the impact of such
contamination would be, we have created a selection of predominantly
foreground sources from two HFF clusters, Abell 2744 and MACS0416, by
determining the $V_{606}-Y_{105}$ vs. $J_{125}$ color magnitude
relation for cluster galaxies and then including sources which lie
within $|\Delta(V_{606}-Y_{105})| < 0.2$ mag of the relation.  Nominal
magnification factors are then calculated for sources as if they were
$z\sim5$ galaxies, sources are segregated into different magnification
bins, and then the surface density of these sources is derived as a
function of the nominal magnification factor.  The result is presented
in the lower right panel of Figure~\ref{fig:surfddd} as the red line.
While this selection likely includes a small number of distant
sources, the slope and shape reflects the impact of contaminating
cluster galaxies.

Clearly, the surface density vs. magnification relationship is
different from the behavior from LFs with a relatively shallow
faint-end slope, e.g., $\gtrsim -1.8$, where the surface density of
sources decreases towards higher magnification factors.  If
contaminants are not eliminated from the intermediate and
high-redshift samples, it could have a substantial impact on the
recovered LF results using lensing clusters.\\

\begin{figure*}
\epsscale{0.7}
\plotone{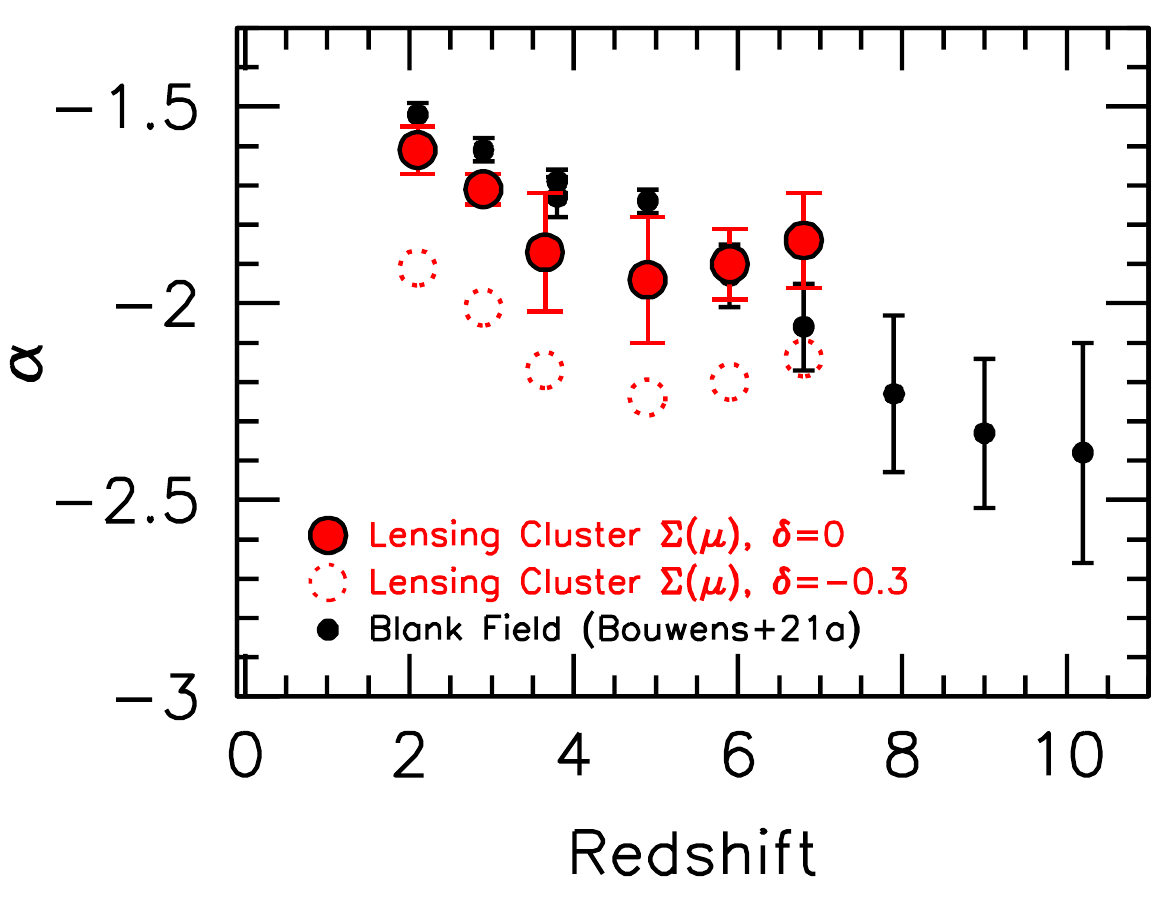}
\caption{Determinations of the faint-end slope $\alpha$ vs. redshift
  (\textit{red filled circles}) from our HFF surface density analysis
  (\S4.2 and \S4.3) and a comparison to blank-field faint end slope
  determinations (\textit{black filled circles}: Bouwens et
  al.\ 2021a).  Error bars are $1\sigma$.  The HFF faint-end slope
  results are derived based on the relationship between the observed
  galaxy surface densities and the magnification factors and assume
  that the selection volumes do not depend on the magnification
  factor, i.e., $\delta = 0$ (as defined in Eq.~\ref{eq:model}); see
  \S4.3.  Also shown are the inferred faint-end slopes $\alpha$
  assuming a magnification-dependent selection efficiency
  $\mu^{-0.3}$, i.e., with $\delta=-0.3$ (\textit{open dotted red
    circles}).  Amazingly, the results we obtain based on the
  magnification-dependent surface densities -- assuming that $S(\mu)$
  is independent of the magnification factor $\mu$, i.e., $\delta=0$
  -- are consistent with blank-field
  determinations.\label{fig:faintend}}
\end{figure*}

In the above paragraphs, we discuss at least five different factors
which can impact the measured surface density of sources
vs. magnification factor.  For some of the described factors, the
impact is similar (e.g., faint-end slope $\alpha$ vs. sizes of faint
sources), so some degeneracies can arise in interpreting the results.

Nevertheless, it is useful for us to construct a fiducial functional
form for interpreting the surface density vs. magnification results.
Based on previous work by Broadhurst (1995), Broadhurst et
al.\ (2005), and Bouwens et al.\ (2017b) and also by the earlier
discussion in this section, the surface density of galaxies in
$z\gtrsim2$ selections can be written as $\mu^{-(2+\alpha)} S(\mu)$.
If we rewrite $S(\mu)$ as a power law with $\mu^{\delta}$, where the
exponent $\delta$ expresses the dependence of the selection efficiency
on the magnification factor:
\begin{equation}
\Sigma(\mu) = \mu^{-(2+\alpha)} \mu^{\delta} = \mu^{\epsilon}
\label{eq:model0}
\end{equation}
where we set $\epsilon$ equal to $\delta-(2+\alpha)$ for simplicity.

As an approximate accounting for the expected predictive power of the
model magnification factors above some value $\mu_{break}$ (\textit{upper
right panel in Figure~\ref{fig:surfddd}}) -- which is not well
established, but seems likely to be anywhere from $\mu\sim10$ to
$\gtrsim$100 -- we can replace $\mu$ with
$\textrm{min}[\mu,\mu_{break}]$ in the above expression.  This results
in our recasting Eq.~\ref{eq:model0} as follows:
\begin{equation}
\Sigma(\mu) = (\textrm{min}[\mu,\mu_{break}])^{\epsilon} = (\textrm{min}[\mu,\mu_{break}])^{\delta-(2+\alpha)}
\label{eq:model}
\end{equation}

Given the challenges uncertainties in the magnification model pose to
characterizing a potential turn-over at lower luminosities (e.g., see
Bouwens et al.\ 2017b, Atek et al.\ 2018) and given that the impact of
one runs counter to the other (\textit{lower center panel of
  Figure~\ref{fig:surfddd}}), we will largely ignore the issue of a
faint-end turn-over for the results we derive in the following
sections.  This issue will nevertheless be revisited in the companion
paper to the present one (R. Bouwens et al. 2022, in prep).

\subsection{Observed Dependence of the Source Surface Densities on the Model Magnification Factors}

Having discussed the expected dependencies on the model magnification
factors, we now proceed to characterize how the surface density of
galaxies in our $z\sim2$-7 selections exhibit the expected traits.  We
focus on the selections for $z\sim2$-7 galaxies because of the much
larger number of sources in those selections.  The $z\sim8$, 9, and 10
samples are too small to map out the trends.

Using the median magnification factor we derive from the parametric
model, we segregate the sources in each of our samples into different
magnification bins, compute the total area available over all six HFF
clusters to identify sources in a given bin of magnification (of width
0.166 and 0.333 dex for our $z\sim2$-3 and $z\sim4$-7 selections,
respectively), and then derive the resultant surface densities.  We
present our results in Figure~\ref{fig:surfd}.

We model the surface densities as a simple power law in the
magnification factor $\mu$, i.e., $\mu^{\epsilon}$, with a break in
the power-law at $\mu_{break} = 30$ and $\mu_{break} = 300$.  In the
former case, we suppose that the model magnification factors retain
their predictive power to 30 and in the latter case to 300.  We
include the latter case to show the behavior in the case that the
models have predictive power beyond values of 30 (as suggested by our
earlier results though it seems unlikely results would be predictive
to values of 300).  In the two cases, we find similar power-law
slopes.  The best-fit $\epsilon$'s derived from our power-law fits are
presented in Table~\ref{tab:slope} and range from $-0.41$ to 0.02.

In the subsections which follow, we will attempt to interpret the
observed trends to derive constraints on the faint-end slope $\alpha$
of the $UV$ LF at $z=2$-7, the dependence of selection efficiency
$S(\mu)$ on the magnification factor (i.e., $\mu^{\delta}$), and the
maximum magnification factor to which the lensing models are
predictive.  

\begin{figure*}
\epsscale{0.95}
\plotone{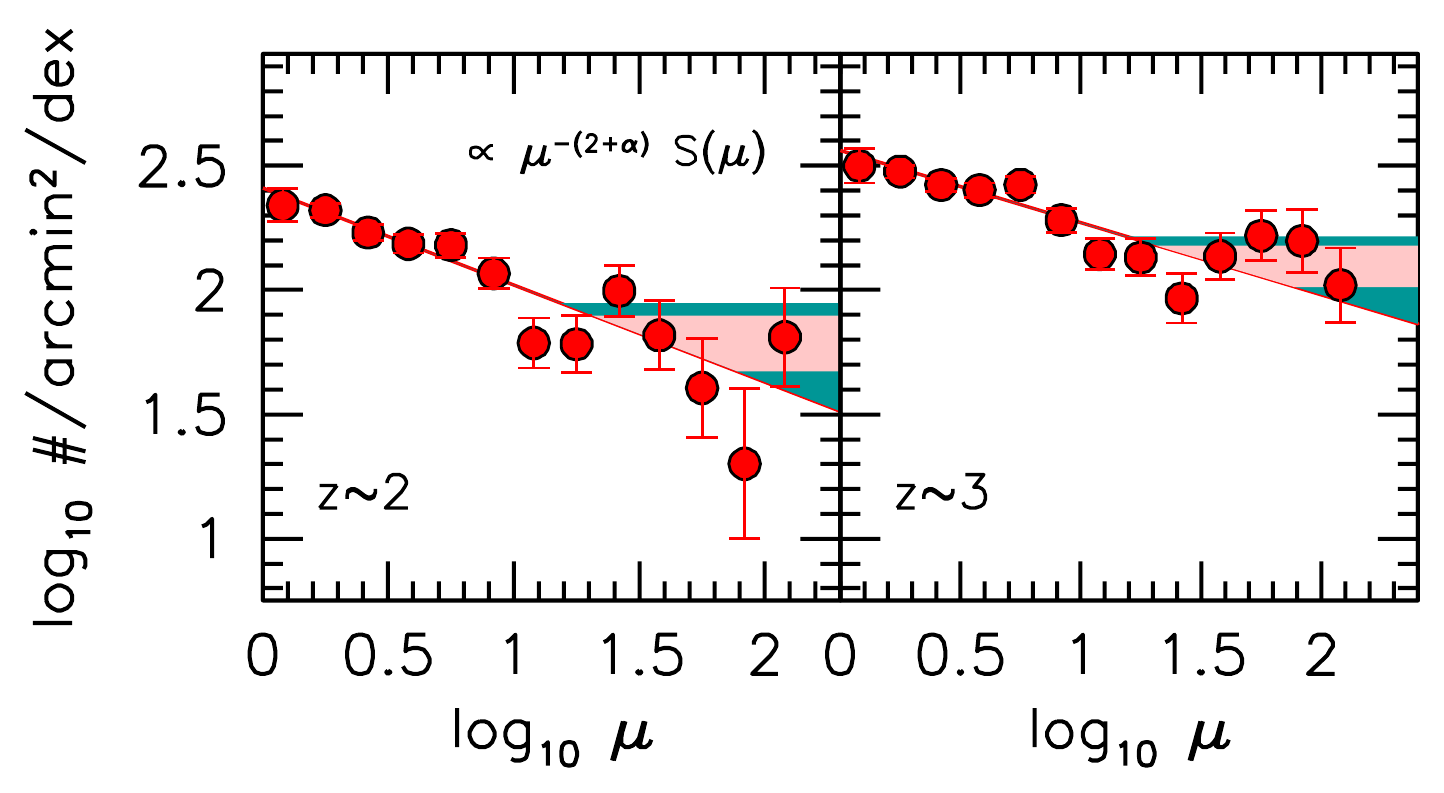}
\caption{Fits to the surface density vs. magnification relation for
  our $z\sim2$ (\textit{left panel}) and $z\sim3$ samples
  (\textit{right panel}: see \S4.5).  Shown with the shaded red and
  green contours are the 68\% and 95\% confidence constraints on the
  relation.  As the upper right panel of Figure~\ref{fig:surfddd}
  illustrates, we would expect the surface density vs. magnification
  trend to maintain a linear form to the magnification factor where
  the models are predictive and then to flatten for even higher
  magnification factors.  These fits suggest that the median of the
  parametric magnification models likely have predictive power to at
  least a magnification factor of 15 and 21 at 95\% and 68\%
  confidence, respectively.  This provides a valuable independent test
  of the reliability of the v4 models to high magnification
  factors.\label{fig:slope_pred}}
\end{figure*}

The challenge we face in attempting to address each of these questions
is that our modeling, i.e., Eq~(\ref{eq:model}), has more free
parameters than constraints, and so some simplifying assumptions will
need to be made.

\subsection{Implications for the Selection Efficiencies}

In our modeling surface density of galaxies as a function of the
magnification $\mu$ using Eq.~\ref{eq:model}, we can simplify the
situation considerably by setting the faint-end slope $\alpha$ in this
equation equal to the faint-end slopes derived from extensive
blank-field studies and then looking for the best-fit values of
$\delta$ and $\mu_{break}$.  Faint-end slopes from blank-field studies
are increasingly well determined and should provide us with a
well-defined reference point for our modeling.  We take the
blank-field slopes from the recent study of Bouwens et al.\ (2021a)
who derive $z\sim2$-9 LF results using a comprehensive set of
blank-field data sets observed with {\it HST}.

The best-fit values of $\delta$ we derive in the $\mu=300$ and
$\mu=30$ cases are 0.09$_{-0.03}^{+0.04}$ and 0.08$\pm$0.04,
respectively.  This is very close to the case $\delta = 0$, where the
selection efficiency shows no dependence on the magnification factor
of sources.  Such a dependence is expected in the case of an
especially steep size-luminosity relation for $z\sim2$-9 galaxies,
i.e., where $r\propto L^{0.5}$ and there is no change in the surface
brightness of sources vs. luminosity or magnification factor at the
faint end of the HFF probes.  Interestingly enough, Bouwens et
al.\ (2017a), Kawamata et al.\ (2018), Bouwens et al.\ (2022), and
Yang et al.\ (2022) all found steep size-luminosity relations at
$z\gtrsim 6$, with radius depending on luminosity as
$L^{0.50\pm0.07}$, $L^{0.46_{-0.09}^{+0.08}}$, $L^{0.40\pm0.04}$, and
$L^{0.48\pm0.08}$.

As important context, a negative $\delta$ would be expected for the
size-luminosity relations derived for brighter field galaxies in the
intermediate or high-redshift universe.  As one example, a $r\propto
L^{0.27}$ relation, as derived by Shibuya et al.\ (2015) for $z=4$-8
galaxies, would result in the surface brightness of galaxies scaling
as the square root of the luminosity.  As a result, lower-luminosity
lensed galaxy samples would feature very low surface brightness
sources, resulting in a decrease in the selection efficiency of
galaxies to higher magnification factors.  Oesch et al.\ (2015) found
$\delta \approx -0.3$ in the simulations they ran.  Note that because
gravitational lensing preserves the surface brightness of sources,
lensing would have little impact in making low surface brightness
galaxies more easily detectable.

Given the above considerations and our formal constraints on $\delta$,
it appears that $\delta$ must be close to 0, and that a significant
fraction of the lower luminosity $z=2$-9 galaxy population have very
small rest-$UV$ sizes.  One potential explanation for a potentially
steeper size-luminosity relation in the rest-$UV$ is the possibility
that only only a small part of a fainter star-forming galaxies may be
experiencing prominent star formation at a time, resulting in a much
smaller apparent physical size (Overzier et al.\ 2008; Ma et
al.\ 2018; Ploeckinger et al.\ 2019; Bouwens et al.\ 2017c, 2021b,
2022).

The present conclusions are very similar to what we previously
concluded from the tests we performed in Bouwens et al.\ (2017a) where
we looked at the distribution of sources as a function of the shear
factor $S$.  This conclusion was bolstered by Bouwens et
al.\ (2017a)'s results on the measured sizes of faint sources after
stacking them along their major shear axes.

Taking $\delta$ to be equal to 0, we tabulate our determinations of
the faint-end slope $\alpha$ of the $UV$ LF in Table~\ref{tab:slope}.
In Figure~\ref{fig:faintend}, we present these faint-end slope
determinations vs. similar determinations of the faint-end slope
$\alpha$ from the comprehensive field studies considered in the
Bouwens et al.\ (2021a) analysis.  The slopes shown in
Figure~\ref{fig:faintend} are for the $\mu_{break}=30$ case.

We can also see how different these faint-end slope determinations are
from the case where the selection efficiency depended on the
magnification factor to the $-0.3$ power, i.e., $\delta=-0.3$, as
Oesch et al.\ (2015) found in the simulations they ran.  If
$\delta=-0.3$, we would derive faint-end slopes that were $\Delta
\alpha \sim 0.3$ steeper than what we derived for our fiducial
$\delta=0$ assumptions.  As an illustration of the impact of such a
change in $\delta$, we show, with the dotted open red circles in
Figure~\ref{fig:faintend}, the faint-end slopes we derive assuming a
$\delta = -0.3$.  There is clearly significant tension between the
$\delta=-0.3$ slopes and those derived from blank-field observations.

\subsection{Predictive Power of Magnification Maps}

As the upper-right panel of Figure~\ref{fig:surfddd} illustrates, the
surface density $\Sigma$ vs. model magnification factor $\mu_{model}$
relation can significantly flatten at high magnification factors when
the models lose their predictive power.  This occurs due to different
regions of the sky being unsuccessfully segregated in terms of their
magnification levels, effectively resulting in a flat relation above
some magnification factor.

As a result of this effect, through a careful modeling of the surface
density vs. magnification factor relation, we can set constraints on
the magnification factor to which the magnification models are
predictive.  Our strongest constraints come from our $z\sim2$ and
$z\sim3$ samples, due to the steepness of the surface density
vs. magnification relation in Figure~\ref{fig:surfd} and significant
statistics at $z\sim2$ and $z\sim3$.  The steepness of the $z\sim2$
and $z\sim3$ relations can be readily contrasted with the flattening
expected at high magnifications when the models lose their predictive
power.

While redshifts $z\sim2$ and $z\sim3$ provide the strongest
constraints, we make use of all of our $z\sim2$-7 results and perform
a simultaneous fit to the $\Sigma$ vs. $\mu_{model}$ relation to find
the value of $\mu_{break}$ which yields the minimum value of $\chi^2$.
We can then determine the range of $\mu_{break}$ which yield
$\chi^2$'s consistent with the best-fit $\mu_{break}$.  We find that
$\mu_{break}$ is constrained to lie in the range 21 to 77 at 68\%
confidence and to be in excess of 15 at 95\% confidence
(Table~\ref{tab:param}).  We illustrate the derived 68\% and 95\%
confidence intervals on the $z\sim2$ and $z\sim3$ surface density
vs. magnification relations in Figure~\ref{fig:slope_pred} using the
light red and cyan shadings, respectively.

As we discussed earlier, there have been efforts to test the
predictive power of lensing model reconstruction efforts by Meneghetti
et al.\ (2017).  Meneghetti et al.\ (2017) produced a full set of mock
observations for two galaxy clusters and obtained reconstructions of
the lensing maps for these clusters based on the mock data sets.
Encouragingly enough, Meneghetti et al.\ (2017) found that many of the
blindly recovered magnification maps were reliable to magnification
factors of at least 30.  This is in excellent agreement with the
results from this section and also earlier in this manuscript
(Figures~\ref{fig:predictind}-\ref{fig:predictlit}) using multiple
independent models of the same cluster.

The constraints in this section, based on the surface density
measurements, have significant utility in evaluating the public
magnification models for the HFF clusters.  They provide us with a
highly independent assessment of both the quality and reliability of
the magnification models.

\begin{deluxetable}{cc}
\tablewidth{7cm}
\tabletypesize{\footnotesize}
\tablecaption{New Constraints Obtained Modeling the 
$\Sigma$ vs. $\mu$ Relation\tablenotemark{a}\label{tab:param}}
\tablehead{
\colhead{Parameter} & \colhead{Constraint}}
\startdata
\multicolumn{2}{c}{Magnification-Dependent Selection Efficiencies}\\
$\delta$ ($\mu=300$) & 0.09$_{-0.03}^{+0.04}$ \\
$\delta$ ($\mu=30$) & 0.08$\pm$0.04 \\
& $\approx$0\tablenotemark{b}\\
\\
\multicolumn{2}{c}{Predictive Power of Magnification Models}\\
$\mu_{break}$ & $<$15 (95\% confidence) \\
             & [21,77] (68\% confidence) 
\enddata
\tablenotetext{a}{Obtained from a joint fit of the surface densities to Eq.(\ref{eq:model}) at all redshifts and using the observed faint-end slopes $\alpha$ from blank-field studies (Bouwens et al.\ 2021a).}
\tablenotetext{b}{The best estimate we obtain for $\delta$ is consistent with the idea that the selection efficiency is independent of the magnification factor.  This implies that very low luminosity galaxies must be very small (Bouwens et al.\ 2017a, 2017c, 2022; Kawamata et al.\ 2018).}
\end{deluxetable}

\subsection{Comparison to Previous Work on the Magnification Bias from the HFFs}

Previously, Leung et al.\ (2018) made use of a deep selection of
$z\geq 4.75$ galaxies behind the six HFF clusters to quantify the
surface density of $z\geq 4.75$ galaxies vs. the magnification factor
$\mu$.  The surface density they derived for sources are highest in
the regions with lower magnification factors and decrease by a factor
of $\sim$5 to magnification factors of $>$50.

This is a slightly stronger dependence than what we find for galaxies
in our $z\sim4$, $z\sim5$, and $z\sim6$ samples, where we only observe
a factor of $\sim$1-2$\times$ decrease in the surface densities from
$\mu$ $\sim$1 to $\sim$50.  It is unclear why Leung et al.\ (2018)
find a stronger dependence than we do, but we note that they only
consider those high magnification regions which show little
contamination from intracluster light in the original images, and so
only have a smaller sample to utilize.

Leung et al.\ (2018) interpret the steeper decline they find in the
surface densities of $z\geq4.75$ galaxies to high magnifications as
providing evidence for a wave $\psi$ dark matter model (Schive et
al.\ 2014) with boson masses spanning the range 0.8$\times$10$^{-22}$
eV to 3.2$\times$10$^{-22}$ eV.  Such a dark matter scenario would
favor a turn-over in the $UV$ LF somewhere between $-15$ and $-16.5$
mag.

While this is an exciting possibility, our surface density results do
not show the same $\sim$5$\times$ decrease at magnification factors
$\gtrsim$10 that Leung et al.\ (2018) find, as also illustrated in
Figure~\ref{fig:surfddd}.  Therefore, at least according to our
measurements, any turn-over in the $UV$ LF -- if it exists -- must
occur at fainter luminosities than $-$16 mag.  Independent constraints
from Castellano et al.\ (2016), Livermore et al.\ (2017), Bouwens et
al.\ (2017b), Ishigaki et al.\ (2018), and Atek et al.\ (2018) are
also suggestive of the same conclusion, constraining a faint-end
turn-over to occur at $\gtrsim$$-$15 mag.

\section{Summary}

In this paper, we have made use of the available {\it HST}
observations over all six HFF cluster fields to assemble a
comprehensive sample of 2534 lensed galaxies over the redshift range
$z\sim2$-9.  Our samples include a total of 765 $z\sim2$, 1176
$z\sim3$, 274 $z\sim6$, 125 $z\sim7$, 51 $z\sim8$, and 16 $z\sim9$
galaxy candidates from the HFF cluster observations.  In addition, 68
and 59 galaxy candidates are found at $z\sim4$ and $z\sim5$,
respectively, using a much more conservative set of selection criteria
(as required to minimize contamination from galaxies in the foreground
clusters: see Figures~\ref{fig:clust_contam} and \ref{fig:purity}).
Including the two $z\sim10$ candidates found by Oesch et al.\ (2018a),
the cumulative sample of lensed $z=2$-10 galaxies includes 2536
distinct sources.

We estimate the luminosity of these sources by utilizing the median
magnification factor from the latest (i.e., v3/v4) HFF lensing models
(Table~\ref{tab:models}).  These models make use of the comprehensive
set of multiple image pairs available from the HFF observations, as
well as a comprehensive set of spectroscopic redshift constraints.

By taking advantage of the availability of multiple independent models
of each HFF cluster and alternatively treating one of the v4
parametric models as the truth, we assess how well the other models,
considered both individually or as a median, are able to predict the
magnification factors seen in the ``true'' model.  We find that
individual v4 magnification models are effective in predicting the
magnification factors in other models to magnifications of $\sim$20-50
(Figure~\ref{fig:predictind}).  When using the median magnification
maps, the models are predictive to magnification factors of
$\gtrsim$40 and in the case of Abell S1063, the predictive range
appears to extend to magnification factors of $\sim$100.

We also assessed the robustness of the published magnification factors
derived for different $z\sim6$-10 samples from the literature (Zitrin
et al.\ 2014; Infante et al.\ 2015; Livermore et al.\ 2017; Ishigaki
et al.\ 2018).  Using the median magnification factors from the v4
parametric models, we found that published magnification factors
showed a significant degree of robustness.  For magnification factors
in excess of 50 and especially 100, the robustness of published
magnification factors was less, with new estimates typically in the
range 40 to 100.

Our tests cast doubt on the robustness of any magnification estimates
in excess of 100, and therefore we take 100 as the maximum fiducial
magnification factor.  Imposing this magnification limit, the lowest
luminosity galaxies that we identify behind the HFF clusters have $UV$
luminosities of $-$12.4 mag (at $z\sim 3$) and $-$12.9 mag (at $z\sim
7$).

We have made use of these large $z\sim2$-9 samples of magnified
galaxies to characterize the dependence of the surface densities on
the magnification factors.  Examining this relationship is
particularly valuable, since this dependence not only provide us with
constraints on the faint-end slope of the LF, and any possible
turnover, but also gives us insight into the sizes of faint $z\geq2$
galaxies and the predictive power of the lensing models.

We find the slope of the surface density vs. magnification relation is
exactly what we would expect if the selection efficiencies showed no
strong dependence on the magnification factor $\mu$.  This can only be
the case if lower luminosity sources are small, as concluded in
Bouwens et al.\ (2017a, 2021b, 2022) and Kawamata et al.\ (2018).  In
the limit of no dependence of the selection efficiency on the
magnification factor, we derive the faint-end slope of the LFs at
$z\sim2$, 3, 4, 5, 6, and 7.  Excellent agreement is found with
faint-end slope determinations from blank field studies (e.g., Bouwens
et al.\ 2021a).  This consistency strongly supports the conclusions we
draw from the surface density vs. magnification results.

In addition, we use the strong systematic dependence of source surface
density on magnification factor to constrain the predictive power of
magnification models.  If the magnification models are not accurate
above a given magnification factor, one would expect the dependence to
immediately flatten above that magnification factor.  Our results
indicate that the median magnification factors from the parametric
models provide reliable results to at least a magnification factor of
21 and 15 (68 and 95\% confidence, respectively).

In a companion paper (R. Bouwens et al.\ 2022, in prep) to the present
analysis, we will be using the current selection of 2534 $z\sim2$-9
galaxies to set constraints on the faint end form of the $UV$ LFs at
$z=2$-9.  This will include a careful characterization not only of how
the faint-end slope of the $UV$ LF evolves from $z\sim9$ to $z\sim2$,
but also to set limits on possible faint-end turn-overs to these same
$UV$ LFs.

\acknowledgements

We thank Gabriel Caminha for kindly providing us with a copy of the
published lensing model from Abell S1063 in Caminha et al. (2016).  We
acknowledge the support of NASA grants HST-AR-13252, HST-GO-13872, and
HST-GO-13792. RB, MS, and BR gratefully acknowledge support from NWO
grants 600.065.140.11N211 (vrij competitie) and TOP grant TOP1.16.057.
RSE acknowledges funding from the European Research Council (ERC)
under the European Union’s Horizon 2020 research and innovation
programme (grant agreement No 669253).  PO acknowledge support from:
the Swiss National Science Foundation through the SNSF Professorship
grant 190079 (PAO, JK). The Cosmic Dawn Center (DAWN) is funded by the
Danish National Research Foundation under grant No. 140.  This work
utilizes gravitational lensing models produced by PIs Bradac,
Natarajan \& Kneib (CATS), Merten \& Zitrin, Sharon, Williams, Keeton,
Bernstein and Diego, and the \textsc{GLAFIC} group. This lens modeling
was partially funded by the {\it HST} Frontier Fields program
conducted by STScI. STScI is operated by the Association of
Universities for Research in Astronomy, Inc. under NASA contract NAS
5-26555. The lens models were obtained from the Mikulski Archive for
Space Telescopes (MAST).

\noindent \textit{Software:} galfit (Peng et al. 2002, 2010), SExtractor
software (Bertin \& Arnouts 1996), mophongo (Labbe et al. 2010a,
2010b, 2013, 2015), EAzY (Brammer et al. 2008), lenstool (Jullo \&
Kneib 2009), GLAFIC (Oguri 2010)


\begin{thebibliography}{} 
\bibitem[Anders et al. (2003)]{2003AAP..401..1063A} Anders, P., \& Fritze-v.~Alvensleben, U.\ 2003, \aap, 401, 1063
\bibitem[Alavi et al.(2014)]{2014ApJ...780..143A} Alavi, A., Siana, B., Richard, J., et al.\ 2014, \apj, 780, 143 
\bibitem[Alavi et al.(2016)]{2016ApJ...832...56A} Alavi, A., Siana, B., Richard, J., et al.\ 2016, \apj, 832, 56 
\bibitem[Atek et al.(2014)]{2014ApJ...786...60A} Atek, H., Richard, J., Kneib, J.-P., et al.\ 2014, \apj, 786, 60 
\bibitem[Atek et al.(2015)]{2015ApJ...800...18A} Atek, H., Richard, J., Kneib, J.-P., et al.\ 2015a, \apj, 800, 18 
\bibitem[Atek et al.(2015)]{2015ApJ...814...69A} Atek, H., Richard, J., Jauzac, M., et al.\ 2015b, \apj, 814, 69 
\bibitem[Atek et al.(2018)]{2018MNRAS.479.5184A} Atek, H., Richard, J., Kneib, J.-P., \& Schaerer, D.\ 2018, \mnras, 479, 5184 
\bibitem[Beckwith et al.(2006)]{2006AJ....132.1729B} Beckwith, S.~V.~W., 
Stiavelli, M., Koekemoer, A.~M., et al.\ 2006, \aj, 132, 1729 
\bibitem[Bertin and Arnouts (1996)]{1996A&AS..117..393B} Bertin, E.\ and 
  Arnouts, S.\ 1996, \aaps, 117, 39
\bibitem[Bhatawdekar et al.(2019)]{2019MNRAS.486.3805B} Bhatawdekar, R., Conselice, C.~J., Margalef-Bentabol, B., et al.\ 2019, \mnras, 486, 3805
\bibitem[Binney et al. (1987)]{binney} Binney, J., \& Tremaine, S.
  1987, Galactic Dynamics, Princeton University Press
\bibitem[Bouwens et al.(2011)]{2011ApJ...737...90B} Bouwens, R.~J., Illingworth, G.~D., Oesch, P.~A., et al.\ 2011, \apj, 737, 90 
\bibitem[Bouwens et al.(2015)]{2015ApJ...803...34B} Bouwens, R.~J., 
Illingworth, G.~D., Oesch, P.~A., et al.\ 2015, \apj, 803, 34
\bibitem[Bouwens et al.(2017)]{2017ApJ...843...41B} Bouwens, R.~J., Illingworth, G.~D., Oesch, P.~A., et al.\ 2017a, \apj, 843, 41 
\bibitem[Bouwens et al.(2017)]{2017ApJ...843..129B} Bouwens, R.~J., Oesch, P.~A., Illingworth, G.~D., Ellis, R.~S., \& Stefanon, M.\ 2017b, \apj, 843, 129
\bibitem[Bouwens et al.(2017)]{2017c} Bouwens, R.~J., Illingworth,
  G.~D., Oesch, P.~A., et al. 2017c, arXiv:1711.02090  
\bibitem[Bouwens et al.(2021)]{2018submit} Bouwens, R.~J., Oesch, P.A., Stefanon, M., et al. 2021a, \aj, 162, 47
\bibitem[Bouwens et al.(2021)]{2021AJ....162..255B} Bouwens, R.~J., Illingworth, G.~D., van Dokkum, P.~G., et al.\ 2021b, \aj, 162, 255
\bibitem[Bouwens et al. (2022)]{2022ApJ...927..81B} Bouwens, R.~J., Illingworth, G.~D., van Dokkum, P.~G., et al.\ 2022, \apj, 927, 81
\bibitem[Boylan-Kolchin et al.(2015)]{2015MNRAS.453.1503B} Boylan-Kolchin, M., Weisz, D.~R., Johnson, B.~D., et al.\ 2015, \mnras, 453, 1503 
\bibitem[Brada{\v c} et al.(2009)]{2009ApJ...706.1201B} Brada{\v c}, M., Treu, T., Applegate, D., et al.\ 2009, \apj, 706, 1201 
\bibitem[Brammer et al.(2008)]{2008ApJ...686.1503B} Brammer, G.~B., van 
  Dokkum, P.~G., \& Coppi, P.\ 2008, \apj, 686, 1503
\bibitem[Broadhurst(1995)]{1995astro.ph.11150B} Broadhurst, T.\ 1995, astro-ph/9511150
\bibitem[Broadhurst et al.(2005)]{2005ApJ...619L.143B} Broadhurst, T., Takada, M., Umetsu, K., et al.\ 2005, \apjl, 619, L143
\bibitem[Bruzual et al. (2003)]{2003..MNRAS..344.1000}  Bruzual, G., \& Charlot, S.\ 2003, \mnras, 344, 1000
\bibitem[Bunker et al.(2004)]{2004MNRAS.355..374B} Bunker, A.~J., Stanway, 
E.~R., Ellis, R.~S., \& McMahon, R.~G.\ 2004, \mnras, 355, 374
\bibitem[Burgasser et al.(2004)]{2004AJ....127.2856B} Burgasser, A.~J., 
McElwain, M.~W., Kirkpatrick, J.~D., et al.\ 2004, \aj, 127, 2856
\bibitem[Caminha et al.(2016)]{2016A&A...587A..80C} Caminha, G.~B., Grillo, C., Rosati, P., et al.\ 2016, \aap, 587, A80
\bibitem[Caminha et al.(2017)]{2017A&A...600A..90C} Caminha, G.~B., Grillo, C., Rosati, P., et al.\ 2017, \aap, 600, A90 
\bibitem[Castellano et al.(2016)]{2016ApJ...823L..40C} Castellano, M., Yue, B., Ferrara, A., et al.\ 2016b, \apjl, 823, L40 
\bibitem[Coe et al.(2015)]{2015ApJ...800...84C} Coe, D., Bradley, L., \& Zitrin, A.\ 2015, \apj, 800, 84 
\bibitem[Coleman et al. (1980)]{1980ApJS..43..393C} Coleman, G.~D., Wu, C.-C., \& Weedman, D.~W.\ 1980, \apjs, 43, 393
\bibitem[Diego et al.(2005)]{2005MNRAS.362.1247D} Diego, J.~M., Sandvik, H.~B., Protopapas, P., et al.\ 2005, \mnras, 362, 1247 
\bibitem[Diego et al.(2007)]{2007MNRAS.375..958D} Diego, J.~M., Tegmark, M., Protopapas, P., \& Sandvik, H.~B.\ 2007, \mnras, 375, 958 
\bibitem[Diego et al.(2015)]{2015MNRAS.447.3130D} Diego, J.~M., Broadhurst, T., Molnar, S.~M., Lam, D., \& Lim, J.\ 2015a, \mnras, 447, 3130
\bibitem[Diego et al.(2015)]{2015MNRAS.451.3920D} Diego, J.~M., Broadhurst, T., Zitrin, A., et al.\ 2015b, \mnras, 451, 3920 
\bibitem[Diego et al.(2018)]{2018MNRAS.473.4279D} Diego, J.~M., Schmidt, K.~B., Broadhurst, T., et al.\ 2018, \mnras, 473, 4279
\bibitem[Dressel et al. (2012)]{Dressel} Dressel, L., et al.\ 2012. “Wide Field Camera 3 Instrument Handbook, Version 5.0” (Baltimore: STScI)
\bibitem[Illingworth et al.(2013)]{2013ApJS..209....6I} Illingworth, G.~D., Magee, D., Oesch, P.~A., et al.\ 2013, \apjs, 209, 6
\bibitem[Infante et al.(2015)]{2015ApJ...815...18I} Infante, L., Zheng, W., Laporte, N., et al.\ 2015, \apj, 815, 18 
\bibitem[Ishigaki et al.(2018)]{2018ApJ...854...73I} Ishigaki, M., Kawamata, R., Ouchi, M., et al.\ 2018, \apj, 854, 73
\bibitem[Jullo et al.(2009)]{2009MNRAS..395..1319J} Jullo, E., \& Kneib, J.-P.\ 2009, \mnras, 395, 1319
\bibitem[Kawamata et al.(2016)]{2016ApJ...819..114K} Kawamata, R., Oguri, M., Ishigaki, M., Shimasaku, K., \& Ouchi, M.\ 2016, \apj, 819, 114 
\bibitem[Kawamata et al.(2018)]{2018ApJ...855....4K} Kawamata, R., Ishigaki, M., Shimasaku, K., et al.\ 2018, \apj, 855, 4
\bibitem[Keeton(2010)]{2010GReGr..42.2151K} Keeton, C.~R.\ 2010, General Relativity and Gravitation, 42, 2151 
\bibitem[Koekemoer et al.(2013)]{2013ApJS..209....3K} Koekemoer, A.~M., Ellis, R.~S., McLure, R.~J., et al.\ 2013, \apjs, 209, 3
\bibitem[Koekemoer et al.(2014)]{2014AAS...22325402K} Koekemoer, A.~M., Avila, R.~J., Hammer, D., et al.\ 2014, American Astronomical Society Meeting Abstracts \#223
\bibitem[Kotulla et al. (2009)]{2009MNRAS..396..462K} Kotulla, R., Fritze, U., Weilbacher, P., \& Anders, P.\ 2009, \mnras, 396, 462  
\bibitem[Kron (1980)]{kron} Kron, R. G. 1980, \apjs, 43, 305
\bibitem[Kuhlen \& Faucher-Gigu{\`e}re(2012)]{2012MNRAS.423..862K}
  Kuhlen, M., \& Faucher-Gigu{\`e}re, C.-A.\ 2012, \mnras, 423, 862
\bibitem[Labb{\'e} et al.(2010)]{2010ApJ...708L..26L} Labb{\'e}, I., 
Gonz{\'a}lez, V., Bouwens, R.~J., et al.\ 2010a, \apjl, 708, L26
\bibitem[Labb{\'e} et al.(2010)]{2010ApJ...716L.103L} Labb{\'e}, I., 
Gonz{\'a}lez, V., Bouwens, R.~J., et al.\ 2010b, \apjl, 716, L103
\bibitem[Labb{\'e} et al.(2013)]{2013ApJ...777L..19L} Labb{\'e}, I., Oesch, 
  P.~A., Bouwens, R.~J., et al.\ 2013, \apjl, 777, L19
\bibitem[Labb{\'e} et al.(2015)]{2015ApJS..221...23L} Labb{\'e}, I., Oesch, P.~A., Illingworth, G.~D., et al.\ 2015, \apjs, 221, 23
\bibitem[Laporte et al.(2016)]{2016ApJ...820...98L} Laporte, N., Infante, L., Troncoso Iribarren, P., et al.\ 2016, \apj, 820, 98 
\bibitem[Leung et al.(2018)]{2018ApJ...862..156L} Leung, E., Broadhurst, T., Lim, J., et al.\ 2018, \apj, 862, 156 
\bibitem[Liesenborgs et al.(2006)]{2006MNRAS.367.1209L} Liesenborgs, J., De Rijcke, S., \& Dejonghe, H.\ 2006, \mnras, 367, 1209 
\bibitem[Limousin et al.(2016)]{2016A&A...588A..99L} Limousin, M., Richard, J., Jullo, E., et al.\ 2016, \aap, 588, A99 
\bibitem[Livermore et al. (2017)]{livermore} Livermore, R., Finkelstein, S., Lotz, J. 2017, ApJ, 835, 113
\bibitem[Lotz et al.(2017)]{2017ApJ...837...97L} Lotz, J.~M., Koekemoer, A., Coe, D., et al.\ 2017, \apj, 837, 97
\bibitem[Ma et al.(2018)]{2018MNRAS.477..219M} Ma, X., Hopkins, P.~F., Boylan-Kolchin, M., et al.\ 2018, \mnras, 477, 219 
\bibitem[Mahler et al.(2018)]{2018MNRAS.473..663M} Mahler, G., Richard, J., Cl{\'e}ment, B., et al.\ 2018, \mnras, 473, 663
\bibitem[Meneghetti et al.(2017)]{2017MNRAS.472.3177M} Meneghetti, M., Natarajan, P., Coe, D., et al.\ 2017, \mnras, 472, 3177 
\bibitem[Merlin et al.(2016)]{2016A&A...590A..30M} Merlin, E., Amor{\'{\i}}n, R., Castellano, M., et al.\ 2016, \aap, 590, A30 
\bibitem[Merten et al.(2015)]{2015ApJ...806....4M} Merten, J., Meneghetti, M., Postman, M., et al.\ 2015, \apj, 806, 4 
\bibitem[Oesch et al.(2015)]{2015ApJ...808..104O} Oesch, P.~A., Bouwens, R.~J., Illingworth, G.~D., et al.\ 2015, \apj, 808, 104 
\bibitem[Oesch et al.(2018)]{2018ApJ...855..105O} Oesch, P.~A., Bouwens, R.~J., Illingworth, G.~D., Labb{\'e}, I., \& Stefanon, M.\ 2018a, \apj, 855, 105 
\bibitem[Oesch et al.(2018)]{2018ApJS..237...12O} Oesch, P.~A., Montes, M., Reddy, N., et al.\ 2018b, \apjs, 237, 12 
\bibitem[Oguri(2010)]{2010PASJ...62.1017O} Oguri, M.\ 2010, \pasj, 62, 1017 
\bibitem[Oke \& Gunn(1983)]{1983ApJ...266..713O} Oke, J.~B., \& Gunn, 
  J.~E.\ 1983, \apj, 266, 713
\bibitem[Overzier et al.(2008)]{2008ApJ...677...37O} Overzier, R.~A., Heckman, T.~M., Kauffmann, G., et al.\ 2008, \apj, 677, 37
\bibitem[Owers et al.(2011)]{2011ApJ...728...27O} Owers, M.~S., Randall, S.~W., Nulsen, P.~E.~J., et al.\ 2011, \apj, 728, 27 
\bibitem[Peng et al.(2002)]{2002AJ....124..266P} Peng, C.~Y., Ho, L.~C., 
Impey, C.~D., \& Rix, H.-W.\ 2002, \aj, 124, 266
\bibitem[Peng et al.(2010)]{2010AJ....139.2097P} Peng, C.~Y., Ho, L.~C., Impey, C.~D., \& Rix, H.-W.\ 2010, \aj, 139, 2097
\bibitem[Ploeckinger et al.(2019)]{2019MNRAS.484.4379P} Ploeckinger, S., Schaye, J., Hacar, A., et al.\ 2019, \mnras, 484, 4379
\bibitem[Priewe et al.(2017)]{2017MNRAS.465.1030P} Priewe, J., Williams, L.~L.~R., Liesenborgs, J., et al.\ 2017, \mnras, 465, 1030
\bibitem[Richard et al.(2014)]{2014MNRAS.444..268R} Richard, J., Jauzac, M., Limousin, M., et al.\ 2014, \mnras, 444, 268 
\bibitem[Robertson et al.(2015)]{2015ApJ...802L..19R} Robertson, B.~E., 
Ellis, R.~S., Furlanetto, S.~R., \& Dunlop, J.~S.\ 2015, \apjl, 802, L19
\bibitem[Schive et al.(2014)]{2014NatPh..10..496S} Schive, H.-Y., Chiueh, T., \& Broadhurst, T.\ 2014, Nature Physics, 10, 496
\bibitem[Schlafly \& Finkbeiner(2011)]{2011ApJ...737..103S} Schlafly, E.~F. \& Finkbeiner, D.~P.\ 2011, \apj, 737, 103
\bibitem[Schmidt et al. (2014)]{2014ApJ..782..L36} Schmidt, K.B., Treu, T., Brammer., G.B., et al. 2014, ApJ, 782, L36
\bibitem[Sebesta et al.(2016)]{2015arXiv150708960S} Sebesta, K., Williams, L.~L.~R., Mohammed, I., Saha, P., \& Liesenborgs, J.\ 2016, \mnras, 461, 2126 
\bibitem[Shibuya et al.(2015)]{2015ApJS..219...15S} Shibuya, T., Ouchi, M., \& Harikane, Y.\ 2015, \apjs, 219, 15
\bibitem[Shipley et al.(2018)]{2018ApJS..235...14S} Shipley, H.~V., Lange-Vagle, D., Marchesini, D., et al.\ 2018, \apjs, 235, 14 
\bibitem[Szalay et al.(1999)]{1999AJ....117...68S} Szalay, A.~S.,
Connolly, A.~J., \& Szokoly, G.~P.\ 1999, \aj, 117, 68
\bibitem[Turner et al.(1984)]{1984ApJ...284....1T} Turner, E.~L., Ostriker, J.~P., \& Gott, J.~R.\ 1984, \apj, 284, 1
\bibitem[Umetsu \& Broadhurst(2008)]{2008ApJ...684..177U} Umetsu, K., \& Broadhurst, T.\ 2008, \apj, 684, 177 
\bibitem[Vanzella et al.(2014)]{2014ApJ...783L..12V} Vanzella, E., Fontana, A., Zitrin, A., et al.\ 2014, \apjl, 783, L12 
\bibitem[Weisz et al.(2014)]{2014ApJ...794L...3W} Weisz, D.~R., Johnson, B.~D., \& Conroy, C.\ 2014, \apjl, 794, L3
\bibitem[Yang et al.(2022)]{2022arXiv220108858Y} Yang, L.,
  Leethochawalit, N., Treu, T., et al.\ 2022, /mnras, submitted,
  arXiv:2201.08858  
\bibitem[Yue et al.(2018)]{2018ApJ...868..115Y} Yue, B., Castellano, M., Ferrara, A., et al.\ 2018, \apj, 868, 115
\bibitem[Zheng et al.(2014)]{2014ApJ...795...93Z} Zheng, W., Shu, X., Moustakas, J., et al.\ 2014, \apj, 795, 93 
\bibitem[Zheng et al.(2017)]{2017ApJ...836..210Z} Zheng, W., Zitrin, A., Infante, L., et al.\ 2017, \apj, 836, 210 
\bibitem[Zitrin et al.(2012)]{2012MNRAS.423.2308Z} Zitrin, A., Broadhurst, T., Bartelmann, M., et al.\ 2012, \mnras, 423, 2308 
\bibitem[Zitrin et al.(2013)]{2013ApJ...762L..30Z} Zitrin, A., Meneghetti, M., Umetsu, K., et al.\ 2013, \apjl, 762, L30 
\bibitem[Zitrin et al.(2014)]{2014ApJ...793L..12Z} Zitrin, A., Zheng, W., Broadhurst, T., et al.\ 2014, \apjl, 793, L12 
\bibitem[Zitrin et al.(2015)]{2015ApJ...801...44Z} Zitrin, A., Fabris, A., Merten, J., et al.\ 2015, \apj, 801, 44 
\end{thebibliography}
\end{document}